\documentclass[10pt]{iopart}

\usepackage{graphicx}
\usepackage{epstopdf}
\usepackage{color}
\usepackage{epsfig}
\usepackage{cite}
\usepackage{amssymb}
\usepackage{amsfonts}

\usepackage{bm}
\usepackage{caption}
\usepackage{subcaption} 

\usepackage{braket}
\usepackage{float}
\usepackage{hyperref}
\usepackage{caption} 
\usepackage{enumerate}
\newcommand{\aver}[1]{\langle#1\rangle}

\begin{document}

\title{Signatures of nonclassical effects in optical tomograms}

\author{B Sharmila$^1$, K Saumitran$^1$, S Lakshmibala$^1$ and V Balakrishnan$^1$}
\address{$^{1}$ Department of Physics, Indian Institute of Technology Madras, Chennai 600 036, India}

\date{\today}

\begin{abstract}
Several nonclassical effects displayed by  wave packets subject to generic nonlinear Hamiltonians can be identified and assessed directly from tomograms without attempting to reconstruct the Wigner function or the density matrix explicitly. 
We have demonstrated this for both single-mode and bipartite systems.  We have shown that a wide spectrum of effects such as the revival phenomena, quadrature squeezing and Hong-Mandel and Hillery type higher-order squeezing in both the single-mode system and  the double-well Bose-Einstein condensate can be obtained from appropriate tomograms in a straightforward manner. 
We have investigated entropic squeezing of the subsystem state of a bipartite system  as it evolves in time, solely from tomograms.
Further we have identified a quantifier of the extent of entanglement between subsystems which can be readily obtained from the tomogram and which mirrors the qualitative behavior of other measures of entanglement such as the subsystem von Neumann entropy and the subsystem linear entropy. The procedures that we have demonstrated can be readily adapted to multimode systems. 
\end{abstract}

\vspace{2pc}

\maketitle

\ioptwocol

\section{Introduction \label{Sec1}}
 Atom optics provides an ideal framework for investigating nonclassical effects such as squeezing of quantum states and revivals of quantum wave packets. Squeezing is intimately related to quantum noise reduction in the measurement of either of two noncommuting observables. 
 In the case of a pair of quadrature observables, squeezing in one quadrature below the standard quantum noise limit is achieved at the expense of the other quadrature observable, consistent with the Heisenberg uncertainty relation.
 Detection of squeezed states of light \cite {slush} provided early proof of this nonclassical effect. Further sensitive experiments on squeezed light have paved the way for coherent control of vacuum squeezing in the gravitational wave detection band \cite{vahlbruch}.

 It is an interesting fact that  in the case of a single-mode radiation field propagating through a Kerr-like medium \cite{sudhsqueezing}, governed by an effective Hamiltonian $ \hbar \chi{a^{\dagger}}^{2}a^{2}$ squeezing in a field quadrature occurs in the neighbourhood of revivals and two-subpacket fractional revivals of the field wave function.  
 Here, $a^{\dagger}$ and $a$ are the photon creation and destruction operators  satisfying the commutation relation $[a, a^{\dagger}] = 1$, and $\chi$ is the third-order nonlinear susceptibility of the medium. 
 	 An initial wave packet  $\ket{\psi(0)}$  governed by a nonlinear Hamiltonian  is said to revive fully  at an instant $T_{rev}$ during its dynamical evolution if the wave packet $\ket{\psi(T_{rev})}$  differs from $\ket{\psi(0)}$ only by an overall phase. Thus at this instant the overlap $|\langle{\psi(0)}|\psi(T_{rev})\rangle|^{2}$  equals unity. While the Hamiltonian governing the evolution of the system  must be nonlinear for  wave packet revivals to occur  this is not a sufficient condition,  and very specific quantum interference between the basis states that comprise the wave packet is necessary for observing the revival phenomenon \cite{sudhjphys}.  However, in one-dimensional systems at least {\it near-revivals} are expected to occur, signaled by  $|\langle{\psi(0)}|\psi(T_{rev})\rangle|^{2}$  {\it approximately} equal to unity \cite{gutschick}. 
 The revival phenomenon  has been observed in several physical systems \cite {robi}. In some of these, revivals {\it periodically} occur at integer multiples of $T_{rev}$. The system governed by the Kerr-like Hamiltonian  $ \hbar \chi{a^{\dagger}}^{2}a^{2}$ is an example.
 
 Under certain circumstances, fractional revivals of the wave packet can occur at specific instants between two successive revivals \cite{aver}. At these instants the initial wave packet becomes superposed copies of itself.  For instance, at the instant of a $k$-subpacket fractional revival ($k$ being a positive non-zero integer) of an initial  coherent state the wave packet comprises $k$ superposed  coherent states  each of amplitude less than that of the  initial state.
A new wrinkle appears in the revival scenario if the Kerr-like Hamiltonian above is modified to have an additional term proportional to  ${a^{\dagger}}^{3}a^{3}$. The delicate balance between the coefficients of the Kerr-like term and this additional term can lead to super-revivals, i.e., revivals in a system with more than one time scale \cite{agarwalbanerji,bluhm}.  These differ both in the qualitative aspects and in the instants of occurrence of revivals (or near-revivals) from  the simpler case governed by the Kerr-like Hamiltonian. Super-revivals have been experimentally detected in systems of  alkali atoms subject to an external field ( see for instance, \cite{suprev_expt}). 
  
While these systems can be effectively  modeled with Hamiltonians which involve only the field operators,  bipartite systems with Hamiltonians in which operators corresponding to both subsystems appear explicitly provide an ideal framework for examining a wider spectrum of nonclassical effects. In these systems even an initial factored product  state of the individual subsystem states evolves in general, into  different entangled states at different instants. The possibility of revivals of the initial state at a subsequent instant of time, the occurrence of analogues of fractional revivals of the initial state,  two-mode squeezing, higher-order squeezing (i.e., squeezing in higher powers of  quadrature observables) and the possibility of squeezing in the neighbourhood of full (or near-revivals) of the initial state etc., become very sensitive to the extent of entanglement and the ratio of the strengths of the nonlinearity and  coupling between the subsystems.  
  
  While several bipartite systems have been investigated both theoretically and experimentally in this regard, an interesting candidate system on which more recently several experiments on nonclassical effects have been performed are Bose-Einstein condensates (BEC)  trapped in a double well \cite{BECrev, BECsq1, BECsq2}. These typically involve  investigations on ultracold atoms confined by mesoscopic traps. The interference between atoms released from these traps is a sensitive probe of number squeezing. Such experiments have paved the way for applications in continuous-variable quantum information and quantum enhanced magnetometry.  

Identifying the state of a system at instants when it displays nonclassical effects during temporal evolution is a challenging problem  due to the fact that in principle {\it all} moments of {\it all} relevant observables are needed for this purpose.  In practice this is impossible and judicious experiments to measure an optimal set of appropriate tractable observables have to be performed. 
To be able to reconstruct a quantum state we therefore need a set of operators ({\it quorum}) whose statistics gives us {\it  tomographically}  complete information about the state. For optical tomography of a single-mode radiation field (i.e.,  a single system) the set of rotated quadrature operators  \cite{VogelRisken, ibort} given by
 \begin{equation}
\label{eqn:quadop}
\mathbb{X}_{\theta} = \frac{1}{\sqrt{2}} (a^\dagger \rme^{\rmi \theta} + a \rme^{-\rmi \theta}),
\end{equation}
with $\theta$ ranging from  $0$ to $\pi$, constitutes a quorum.  The tomogram $w(X_\theta, \theta)$ of a state with density matrix $\rho$ is then given by \cite{VogelRisken, LvovskyRaymer}
\begin{equation}
\label{eqn:tomogdef}
w(X_\theta, \theta) = \bra{X_\theta, \theta} \rho \ket{X_\theta, \theta}.
\end{equation}
Here $\ket{X_\theta, \theta}$ is the eigenvector of the quadrature operator $\mathbb{X}_{\theta}$ with eigenvalue $X_\theta$, and is given by \cite{barnett}
\begin{eqnarray}
\nonumber \ket{X_\theta, \theta} =\frac{1}{\pi^{1/4}} \exp\left(-\:\frac{X_{\theta}^{2}}{2} \right. &-\:\frac{1}{2} \,\rme^{\rmi 2 \theta} a^{\dagger 2} \\
& \left. + \sqrt{2}\, \rme^{\rmi \theta} X_{\theta} a^{\dagger} \right)\ket{0},
\end{eqnarray}
where $\ket{0}$ is the zero-photon state.
Clearly, $\theta=0 \, (\pi/2)$ corresponds to the position (momentum) quadrature.
 Essentially the tomogram is a collection of probability distributions corresponding to the quadrature operators and  for every $\theta$ it satisfies  
\begin{equation}
\int_{-\infty}^{\infty}  \rmd X_{\theta} \,  w(X_\theta, \theta) =  1.
\end{equation}

These ideas can be extended in a straightforward manner to multimode systems by introducing tomographic variables $(X_{\theta_{i}}, \theta_{i})$ for the $i$th subsystem of the multipartite system and defining quadrature operators corresponding to each of these pairs.
Correspondingly, we have 
\begin{eqnarray}
\nonumber \int_{-\infty}^{\infty}  \rmd X_{\theta_{1}} \int_{-\infty}^{\infty}  \rmd X_{\theta_{2}} &.......\int_{-\infty}^{\infty}  \rmd X_{\theta_{i}}....\\ &w(X_{\theta_{1}}, \theta_{1}, X_{\theta_{2}}, \theta_{2},........) =1.
\end{eqnarray}
(Here, $\theta_{i}$ are constants and the above equation holds for all values of $\theta_{i}$).

Obtaining the relevant tomograms  from experiments is the first step in a somewhat cumbersome procedure for reconstructing  the density matrix or the Wigner quasi-probability distribution, and hence the state.  Homodyne measurements of the quorum of rotated quadrature operators  are made on an ensemble of identical copies of the system and this quadrature histogram (the tomogram) is used for state reconstruction.
 Various stages in this procedure can be carried out only approximately and maximum likelihood estimates of the quantum state starting from the tomograms obtained experimentally, are inherently error-prone. It would therefore be very useful to `read-off' as much information about a state  directly from the tomogram itself. In particular, identifying signatures of nonclassical effects through simple manipulations of the relevant tomograms becomes an interesting and important exercise. As a step  in this direction,  a  single-mode radiation field
whose dynamics is governed by the Kerr-like Hamiltonian has been considered, and its tomograms at instants of fractional revivals obtained from the corresponding density operators (which are theoretically straightforward to calculate) and examined for signatures of nonclassical effects.  It has been shown that at the instant of a k-subpacket fractional revival there are k `strands' in the corresponding tomogram~\cite{sudhrohithrev}.   
Further, tomograms of entangled states carry a considerable amount of information about the full system. Signatures of entanglement  in the tomogram obtained  at the output port of a  quantum beamsplitter whose input is  a factored product of a cat state and the vacuum state have also been investigated \cite{sudhrohithbs}. 
While these investigations  employ the `inverse procedure' of starting with the known state to obtain the tomogram,  such studies on `known' systems are necessary for acquiring knowledge about  how tomographic patterns mirror nonclassical effects,  so that  state-reconstruction procedures can be avoided when examining nonclassical effects in new systems. 

Many more detailed investigations  need to be carried out before a good understanding can be obtained on how to infer nonclassical properties of a state from its tomogram.  In single-mode systems in which super-revivals occur, straightforward correlations between tomograms and fractional revivals  are in general not expected.  Since several physical systems are governed  by Hamiltonians with terms that have cubic and higher powers of the photon number operators it becomes both  important and relevant to  examine the connection between the nature of the tomogram and the revival phenomena in this case. 
 Further, even in  single-mode systems  and simple bipartite systems with continuous variables, squeezing and  higher-order squeezing properties have not been investigated by merely exploiting tomograms. 

Another aspect concerns the fact that  in a bipartite system we can define a probability distribution $w(X_{\theta_{i}}, \theta_{i})$ ($i = 1,2$) corresponding to the $i$th subsystem and correspondingly an information entropy  $S({\theta_{i}})$  which is the quantum analogue of the  classical Shannon entropy. This is  of the form 
$- \int_{-\infty}^{\infty}  \rmd X_{\theta_{i}} \,w_{i} \, \log w_{i}$.  The quantity $ w_{1} = w(X_{\theta_{1}}, \theta_{1})$ for instance,  is obtained by fixing $\theta_{2}$ and integrating over $X_{\theta_{2}}$.  For any fixed value of $\theta_{1}$,   interesting properties of the entropy have been discussed \cite{orlowski}.  For example,  $S(\theta_{1})  + S(\theta_{1} + \pi/2) \geq  (1 + \log {\pi}) $.  Clearly, if $\theta_{1}$ is set equal to zero we obtain a `position-momentum' entropic uncertainty relation. For Gaussian distributions, as happens for coherent states and the vacuum, this entropic sum satisfies the equality with both entropies being equal to  $(1/2) (1 + \log {\pi})$. There is an upper bound on the entropy. Corresponding to the $x$-quadrature, the entropy $S(x)$ in a state satisfies
  $S(x)  \leq (1/2)[ 1 + \log {\pi} + \log(2 (\Delta x)^{2}) ]$, where $ (\Delta x)^{2}$ is the variance  in $x$ in that state. (Since the variance in the coherent state or in the vacuum is equal to $1/2$, it does not contribute to the information entropy of these states). Similar statements hold for the p-quadrature. As a consequence of this connection between the variance and the information entropy,  a state can exhibit entropic squeezing if in either quadrature the entropy is less than  $(1/2) (1 + \log {\pi})$. This would evidently hold for states whose appropriate quadrature variance is between $0$ and $1/2$. The entropic uncertainty relation guarantees that a state cannot exhibit squeezing in both quadratures. It is evident that as the state of a bipartite system evolves in time the information entropy  will  also change. An interesting and useful exercise  would be to estimate this entropy directly from the subsystem's tomogram corresponding to either quadrature. We will carry out this exercise for a double-well Bose-Einstein condensate in a later section.
  Another property of the bipartite system  that we infer from the tomogram is the extent of entanglement between the two subsystems. This is usually quantified by the subsystem von Neumann entropy (SVNE) or the subsystem linear entropy (SLE)  which are equal to zero for unentangled states. Both the SVNE and the SLE will have the same qualitative features. The SVNE corresponding to  subsystem $A$ is defined as
\begin{equation}
S_{A}=-\mathrm{Tr}(\rho_{A} \, \log \, \rho_{A})=-\sum_{i}\lambda_{i} \, \log \, \lambda_{i}.
\label{eqn:SVNE}
\end{equation}
Here $\lbrace \lambda_{i} \rbrace$ are the eigenvalues of the reduced density matrix $\rho_{A}$ of the subsystem $A$. The SLE corresponding to subsystem $A$ is given by
\begin{equation}
\delta_{A}=1-\mathrm{Tr}(\rho_{A}^{2})=1-\sum_{i}\lambda_{i}^{2}.
\label{eqn:SLE}
\end{equation}
$\rho_{A}$ itself is obtained from the density matrix $\rho$ of the full system by taking the trace over the basis states of $B$.  Equivalently, we can examine the SVNE and the SLE of subsystem $B$, for which similar definitions hold.
We obtain  a quantifier of entanglement from the tomogram which closely mimics the behavior of both these entropies at every instant as the BEC evolves in time. We therefore establish that an entanglement measure can be directly obtained from the tomogram without reconstructing the state (equivalently the density matrix).

The plan of the paper is as follows: 
 In the next section we examine the tomograms of  a  single-mode  radiation field with an effective  Hamiltonian of the form $(\hbar \chi_{1} {a^{\dagger}}^{2} a^{2} + \hbar \chi_{2} {a^{\dagger}}^{3} a^{3})$, with $\chi_{ 1}$ and $\chi_{2}$ being constants with appropriate physical dimensions. We directly obtain properties of full and fractional revivals, amplitude squeezing and higher-order squeezing from the instantaneous tomograms without indulging in state-reconstruction procedures. Since these effects are sensitive to the precise initial state considered, we have carried out our investigations  with initial states $\ket{\alpha}$ ($\alpha \in \mathbb{C}$) that exhibit ideal coherence  and with states that depart from coherence in a quantifiable manner.  
 The latter are 1-photon-added coherent states $\ket{\alpha,1}$ (PACS) and can be obtained from  $\ket{\alpha}$ by applying $a^{\dagger}$ on it and normalizing the state.  This state has been identified experimentally  \cite{zavatta}  using quantum state tomography and is therefore an ideal candidate for our purpose. (In general an $m$-photon-added coherent state (m-PACS) is obtained by repeated application of $a^{\dagger}$ $m$ times on $\ket{\alpha}$ and appropriately normalizing it). 
 
 In Section 3, we extend our investigations to a double-well BEC with interacting atomic species that have Kerr-like nonlinearities. 
 We have identified and assessed the manner in which the revival phenomena and two-mode amplitude squeezing properties are manifested in this system for initial states which are factored products of the states of the individual subsystems. These are combinations of $\ket{\alpha}$ 
 and $\ket{\alpha,1}$  (the latter quantifying departure from macroscopic coherence of the initial condensate), with the understanding that the operators and number states refer in this case to the atomic species. 

Further, the extent of entropic squeezing of the  condensate  in one of the wells  as the system evolves in time,  has been examined and the role played by the precise form of the initial state of the BEC in this context investigated directly from the tomograms at various instants.
We have also  identified a quantifier of the extent of entanglement between the two subsystems of condensates which can be inferred from the tomogram and compared its temporal behavior with that of the SVNE and the SLE.

We conclude with brief comments on the results of our investigations.

\section{Single-mode system: A tomographic approach \label{Sec2}}
\subsection{Revivals and fractional revivals}

 In this section we investigate the manner in which full and fractional revivals are mirrored in tomograms as a single-mode system with Hamiltonian $H =( \hbar \chi_{1} {a^{\dagger}}^{2} a^{2} + \hbar \chi_{2} {a^{\dagger}}^{3} a^{3})$,
 evolves in time. Recall that $a$ and $a^{\dagger}$ are respectively photon destruction and creation operators. In what follows we set $\hbar = 1$ for convenience.

Since we wish to investigate how each of the  two terms in $H$ affects the tomograms, we first consider a system with effective Hamiltonian $H^{'} =  \chi {a^{\dagger}}^{3} a^{3} = \chi N (N-1) (N - 2)$, where $N = a^{\dagger}a$ and $\chi$ is a constant.  
 The tomogram $w(X_{\theta}, \theta)$ of a single-mode system can be seen to have the symmetry property 
\begin{equation}
w(X_\theta, \theta + \pi ) = w(X_\theta, \theta),
\end{equation}
and hence information for $0 \leq \theta < \pi$ is sufficient in principle, to reconstruct the state. However we choose the full range  $0 \leq \theta < 2\pi$  to help visualize  various features of the tomogram better. A convenient expression for $w(X_\theta, \theta)$  has been derived in terms of Hermite polynomials  by  realizing that
\begin{equation} \label{eqn:phaseshift}
\ket{X_\theta, \theta} = \rme^{\rmi \theta a^{\dagger} a} \ket{X},
\end{equation}
 where $\ket{X}$ is the eigenstate of the position operator. It has been shown \cite{MankoFockStates} that as a consequence  of  \eref{eqn:tomogdef} and \eref{eqn:phaseshift} the tomogram of a normalized  pure state $\ket{\psi}$, which can be expanded in the photon number basis as $\sum_{n=0}^{\infty} c_n \ket{n}$, is given by
\begin{equation}
\label{eqn:tomogfock}
w(X_\theta, \theta) = \frac{\rme^{-X_\theta^2}}{\sqrt{\pi}} \left| \sum_{n=0}^{\infty} \frac{c_n \rme^{-\rmi n \theta} }{\sqrt{n!} 2^{\frac{n}{2}} } H_n(X_\theta)  \right|^{2} ,
\end{equation}
where  $H_n(X_\theta)$ are Hermite polynomials. We will use this expression as the time-dependence is reflected only in the coefficients  $c_{n}$ of the  number  basis thus facilitating  numerical computations.  

We use the method of `strand-counting' in the tomogram  to study revival patterns. For a Kerr-like Hamiltonian it has been shown that the number of strands in a tomogram is equal to the number of subpackets at  instants of fractional revivals \cite{sudhrohithrev}.  A limitation in this method is that  individual  strands in the tomogram corresponding to a $k$-subpacket fractional revival will not be distinct for $k$  sufficiently large (say 5 or more) due to quantum interference effects. However, to understand the broad features of revival phenomena it suffices to employ   this  procedure  without resorting to state-reconstruction methods.  We corroborate our numerical findings with  analytical explanations for the revival patterns that we observe. 

\begin{figure*}
	\centering    
	\includegraphics[width=0.3\textwidth]{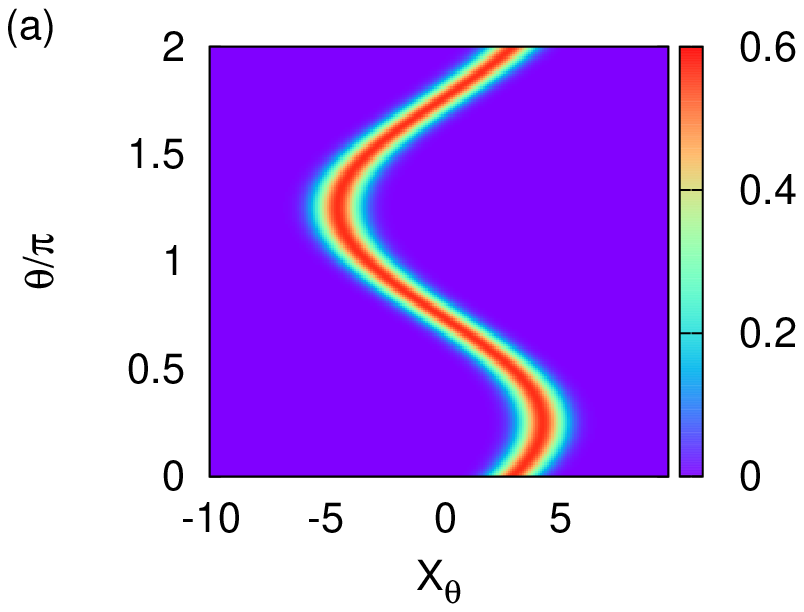}
	\includegraphics[width=0.3\textwidth]{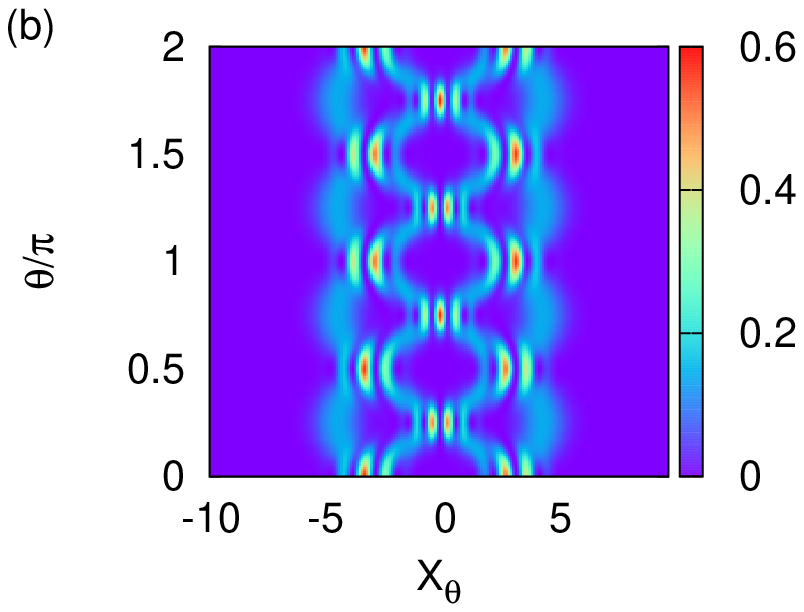}
	\includegraphics[width=0.3\textwidth]{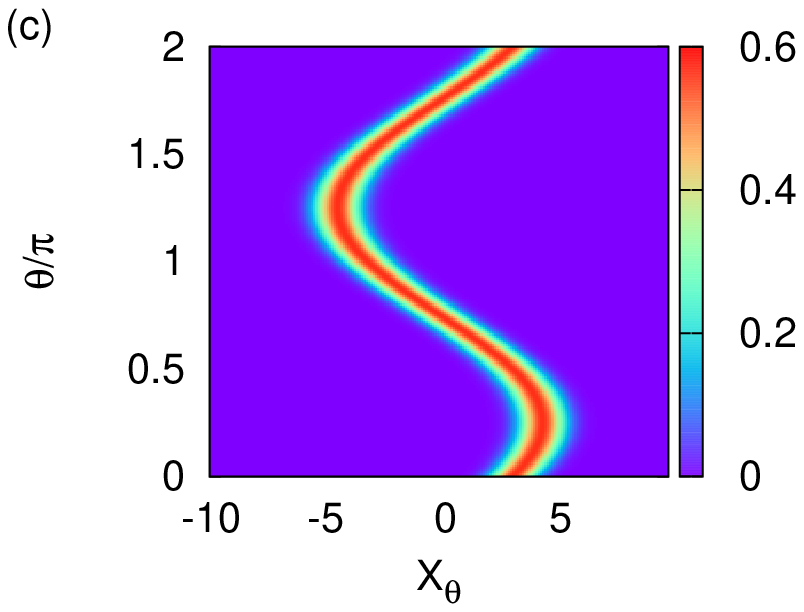}
	\includegraphics[width=0.3\textwidth]{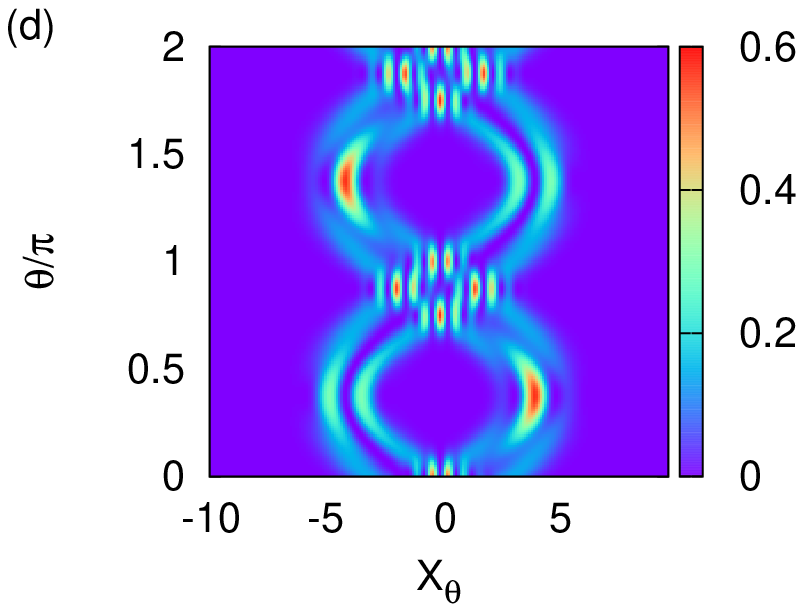}
	\includegraphics[width=0.3\textwidth]{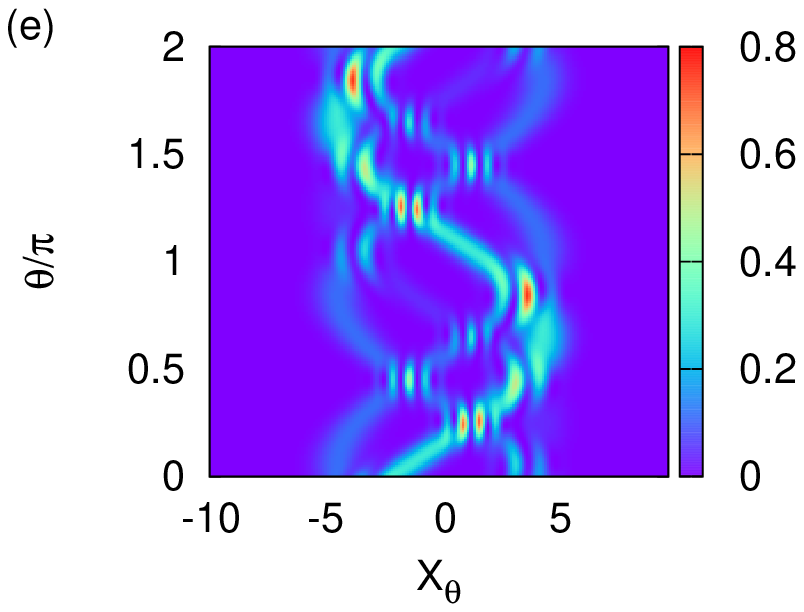}
	\includegraphics[width=0.3\textwidth]{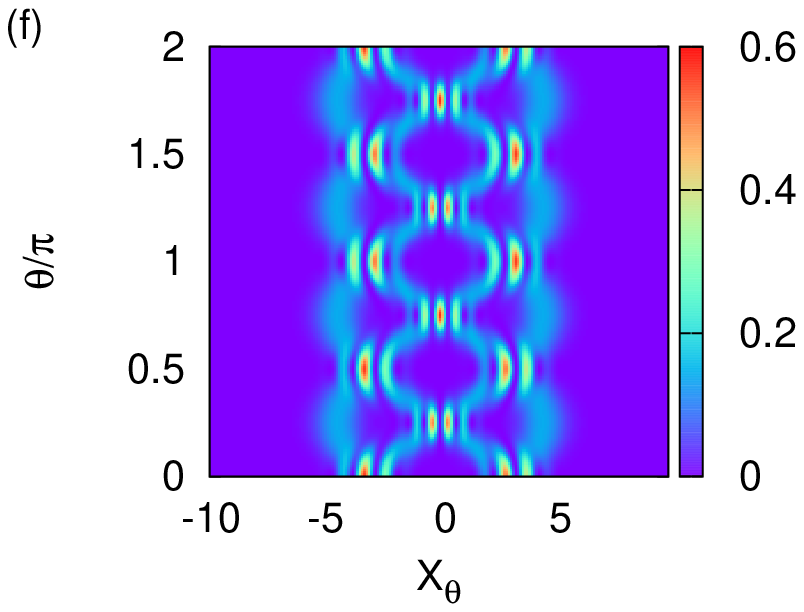}
	\includegraphics[width=0.3\textwidth]{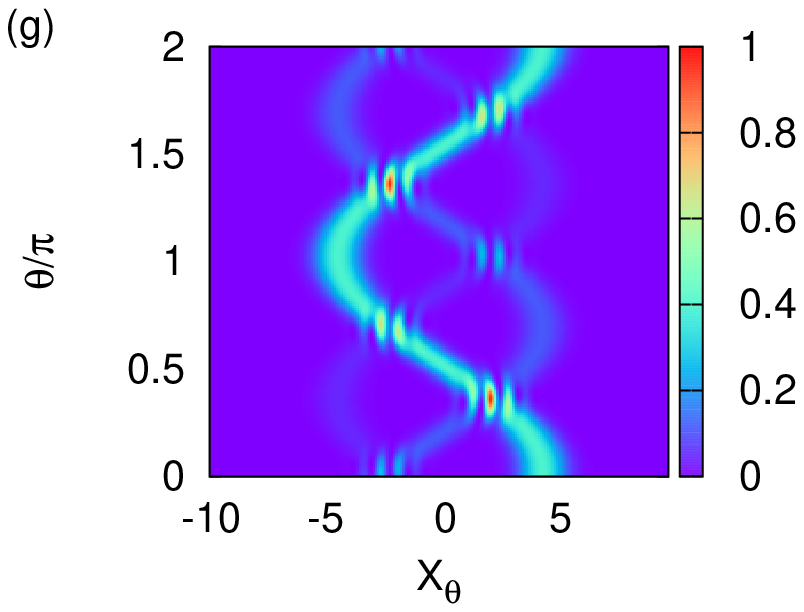}
	\includegraphics[width=0.3\textwidth]{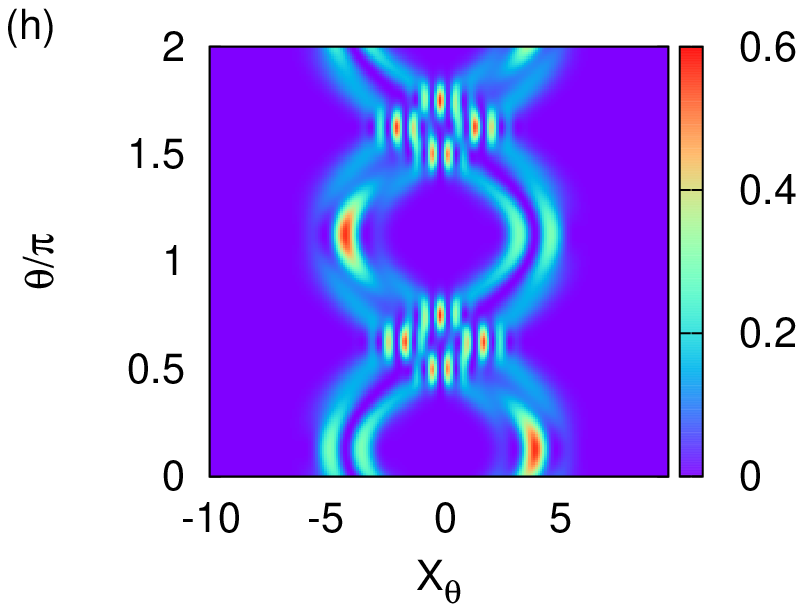}
	\includegraphics[width=0.3\textwidth]{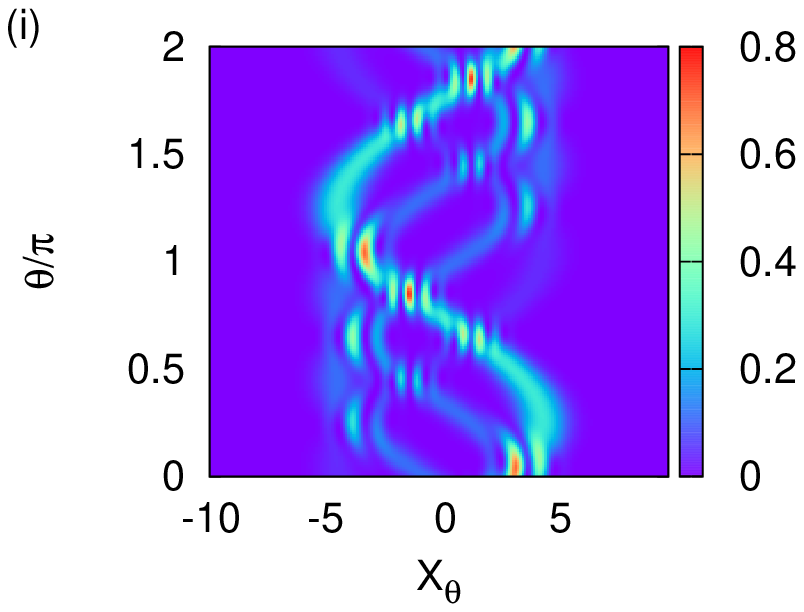}
	\caption{Tomograms of an initial CS with $\alpha=\sqrt{10}e^{i\pi/4}$ for a cubic Hamiltonian at  instants (a) $0$ and $T_{rev}$, (b) $T_{rev}/2$, (c) $T_{rev}/3$, (d) $T_{rev}/4$, (e) $T_{rev}/5$, (f) $T_{rev}/6$, (g) $T_{rev}/9$, (h) $T_{rev}/12$, and (i) $T_{rev}/15$.}
	\label{fig:tomogrevivalcubic}
\end{figure*}

In the system with Hamiltonian $H^{'}$ it can be easily seen that an initial CS or PACS revives fully at instants $ T_{rev} = \pi/\chi$. Hence we examine tomogram patterns at $T_{rev}$ and  at fractional revival times, i.e., at instants
\begin{equation}
t  =  \frac{\pi}{l \chi} = \frac{T_{rev}}{l},
\end{equation}
where $l$ is a positive  integer. The tomograms of an initial CS  corresponding to specific fractional revivals  are shown in figures~\ref{fig:tomogrevivalcubic}(a)-(i). We observe that at both $t = 0$ (equivalently $T_{rev}$) and $T_{rev}/3$ the tomograms look similar.
 Again at  instants $T_{rev}/2$ and $T_{rev}/6$  the tomograms are similar.

These features follow from the properties of the unitary time evolution operator corresponding to this system. It is known for instance that for the simpler system with Hamiltonian ${a^{\dagger}}^{2} a^{2}$ 
the number of subpackets  $p$ of a wave packet at an instant of fractional revival $T_{rev}/p$ is a consequence of the periodicity of the unitary time evolution operator which can be Fourier decomposed at that instant  in the form 
\begin{equation}
U\left(T_{rev}/p\right) = \sum_{m=0}^{p-1} f_m \exp\left(  -\frac{2 \pi \rmi m}{p} N \right),
\end{equation}
where $f_m $ is a Fourier coefficient.  As a consequence, an initial state $\ket{\alpha}$ evolves to a superposition of $p$ coherent states at that instant
\cite{TaraAgarwal}.

While the system at hand is more complicated, the full revival at $T_{rev}/3$  is  a simple consequence of the fact that $n(n-1)(n-2)/3$ is even $\forall\, n\epsilon\mathbb{N}$. Here, $N \ket{n}  = n \ket{n}$, with $ \{\ket{n} \}$ denoting the photon number basis. Hence corresponding to  an initial state $\ket{\psi(0)} = \sum_{m=0}^{\infty} c_{n} \ket{n}$, the state at instant $T_{rev}/3$ is
\begin{eqnarray}
\nonumber\ket{\psi(T_{rev}/3)} &= U(T_{rev}/3) \ket{\psi(0)} \\
\nonumber &= \sum_{n=0}^{\infty} \rme^{-\rmi\pi n(n-1)(n-2)/3} c_n \ket{n} \\
\nonumber &= \ket{\psi(0)}.
\end{eqnarray}

An analysis of the properties of the time evolution operator would, in principle,  explain the appearance of a specific number of strands in the tomogram at different instants $T_{rev}/l$. However accounting for the number of strands in a tomogram is not always straightforward in this case, in contrast to the Kerr-like system.  
 
\begin{figure}
	\centering
	\includegraphics[width=0.4\textwidth]{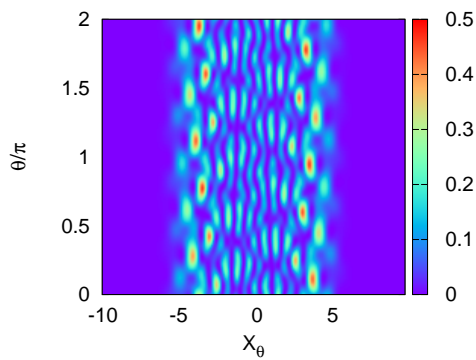}
	\caption{Tomogram of an initial CS with  $\alpha=\sqrt{10}e^{i\pi/4}$, $\chi _{1}= 2.0$ and $\chi_{2}= \sqrt{2}$ at $t=\pi / \chi_{2}$.}
	\label{fig:tomogcsquadcubirrat}
\end{figure}

 We are now in a position to investigate tomogram patterns for the full Hamiltonian 
\begin{equation}
H =  (\chi_{1} a^{\dagger 2} a^2 + \chi_{2} a^{\dagger 3} a^3).
\end{equation}
 In this case, 
\begin{equation}
 T_{rev} = \pi \,  \mathrm {LCM}\left(\frac{1}{\chi_{1}}, \frac{1}{\chi_{2}}\right).
\end{equation}
 As before, we proceed to examine tomogram patterns at different instants of time. 
 If the ratio $\chi_{1}/\chi_{2}$ is irrational,  revivals are absent and the generic tomogram at any instant is blurred.  This is illustrated in \fref{fig:tomogcsquadcubirrat}
 for an initial CS with $\alpha = \sqrt{10} \exp(i \pi /4)$  with $\chi_{1} = 2$, and $\chi_{2} = \sqrt{2}$ at $t = \pi/\chi_{2}$.

\begin{figure}
	\centering    
	\includegraphics[width=0.4\textwidth]{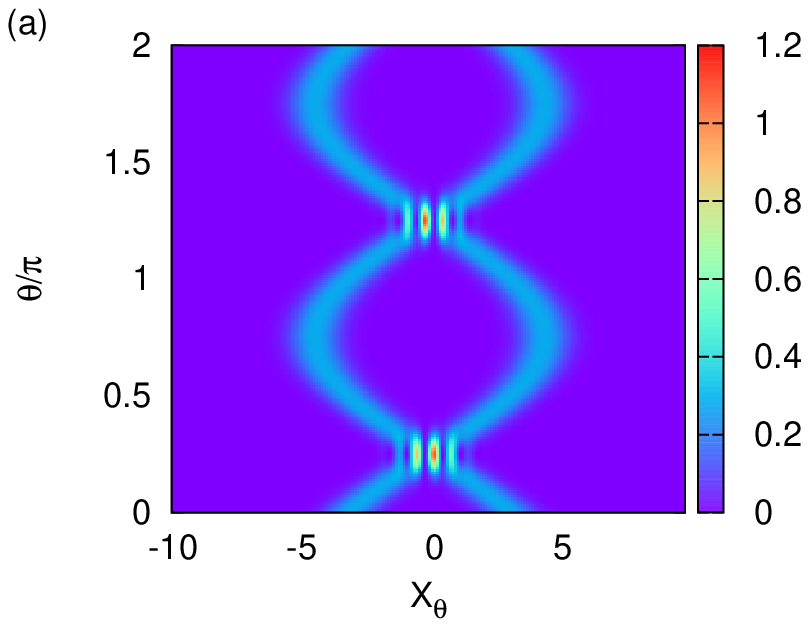}
	\includegraphics[width=0.4\textwidth]{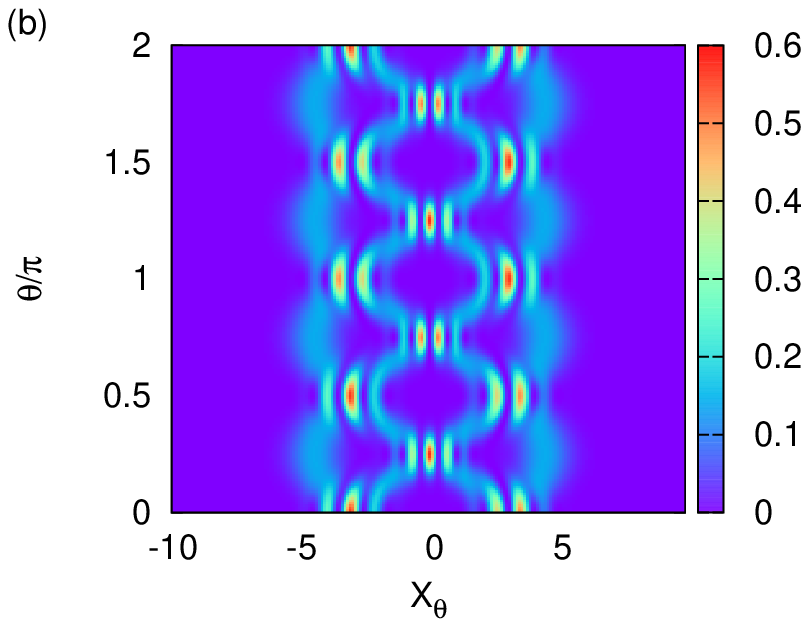}
	\caption{Tomograms of an initial CS with $\alpha=\sqrt{10}e^{i\pi/4}$ at 
$ t = \frac{T_{rev}}{2}$ for (a) $\chi_{1}=1$ and $\chi_{2}=2$ and (b) $\chi_{1}=1$ and $\chi_{2}=3$.}
	\label{fig:tomogrevivalquadcubtrev2}
\end{figure}

\begin{figure}
	\centering    
	\includegraphics[width=0.4\textwidth]{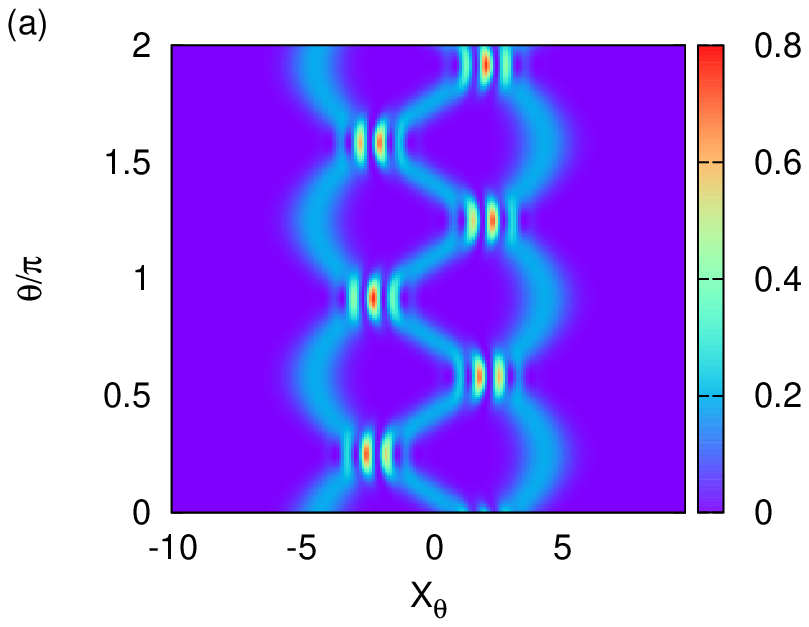}
	\includegraphics[width=0.4\textwidth]{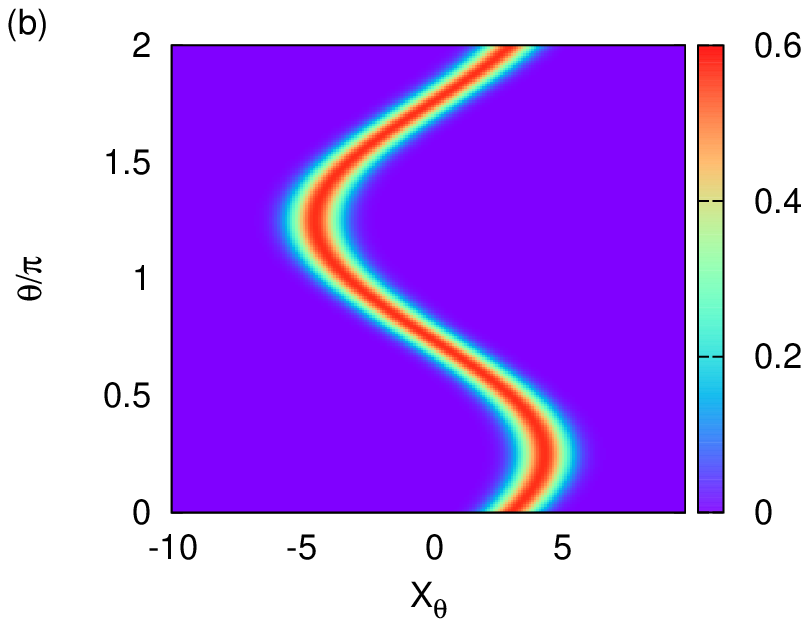}
	\caption{Tomograms of an initial CS with $\alpha=\sqrt{10}e^{i\pi/4}$ at 
$t = \frac{T_{rev}}{3}$ for (a) $\chi_{1}=1$ and $\chi_{2}=2$ and (b) $\chi_{1}=3$ and $\chi_{2}=4$.}
	\label{fig:tomogrevivalquadcubtrev3}
\end{figure}

\begin{figure*}
	\centering    
	\includegraphics[width=0.32\textwidth]{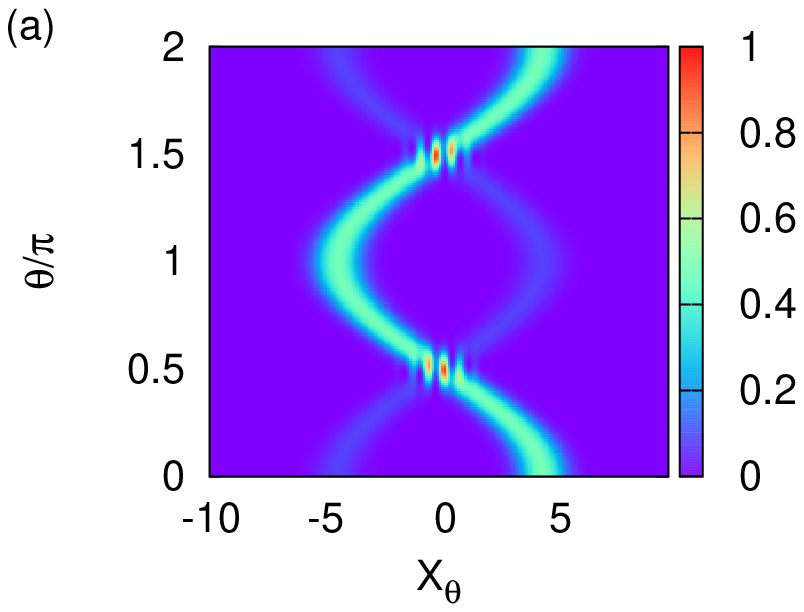}
	\includegraphics[width=0.32\textwidth]{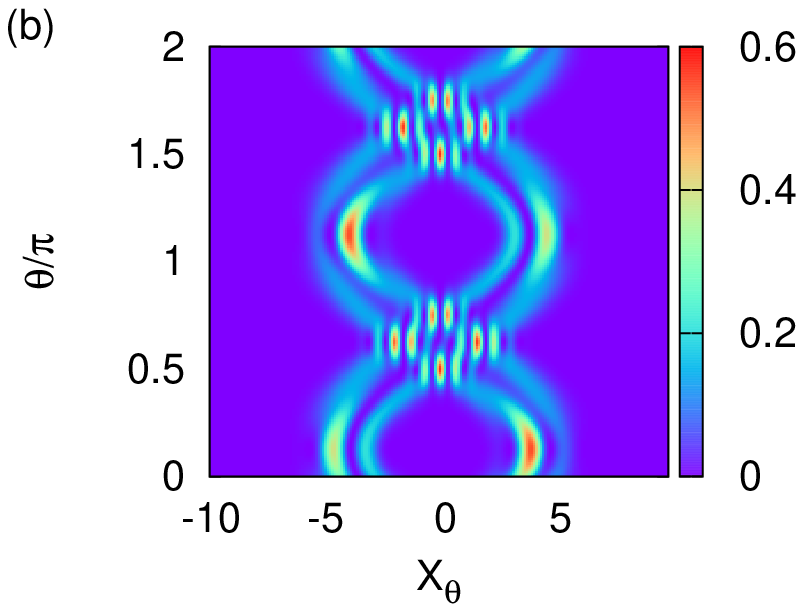}
	\includegraphics[width=0.32\textwidth]{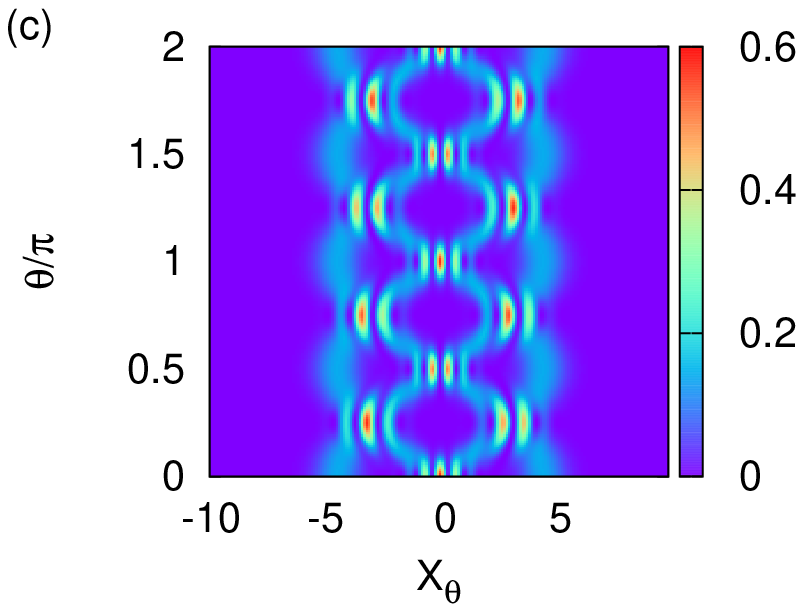}
	\caption{Tomograms of an initial CS with $\alpha=\sqrt{10}e^{i\pi/4}$ 
  at $t=\frac{T_{rev}}{4}$ for (a) $\chi_{1}=1$ and $\chi_{2}=2$, (b)  $\chi_{1}=1$ and $\chi_{2}=3$, and (c) $\chi_{1}=1$ and $\chi_{2}=4/3$.}
	\label{fig:tomogrevivalquadcubtrev4}
\end{figure*}

For rational  $\chi_{1}/\chi_{2}$  revivals and fractional revivals are seen.  Fractional revivals  occur at instants $T_{rev}/l$ as before, but the corresponding tomogram patterns  are sensitive to the ratio $\chi_{1}/\chi_{2}$.  
  We expect that  for a given $l$ the tomogram will have $l$ strands as a consequence of the Kerr-like term in $H$.  While this is one possibility,  the effect of the cubic term in $H$ allows for the possibility of other tomogram patterns.  We illustrate this for an initial CS with Hamiltonian $H$  and 
$\alpha = \sqrt{10} \exp(i \pi /4)$  in  figures \ref{fig:tomogrevivalquadcubtrev2}, \ref{fig:tomogrevivalquadcubtrev3}, and \ref{fig:tomogrevivalquadcubtrev4}.  
 At $t = T_{rev}/2$,  apart from the two-strand tomogram for $\chi_{1} = 1$ and $\chi_{2} = 2$, 
 one of the other possibilities is a four-strand tomogram for $\chi_{1} = 1$ and $\chi_{2} = 3$ (\fref{fig:tomogrevivalquadcubtrev2}). Similarly at $t = T_{rev}/3$, the tomogram has three strands for $\chi_{1} = 1$ and $\chi_{2} = 2$  and a single strand  similar to the tomogram of a CS for 
$\chi_{1} = 3$ and $\chi_{2} = 4$ (\fref{fig:tomogrevivalquadcubtrev3}). At $t = T_{rev}/4$ three specimen tomograms which are distinctly different from each other are shown in figures \ref{fig:tomogrevivalquadcubtrev4} (a)-(c) for $\chi_{1} = 1$ and $\chi_{2} = 2$,  $\chi_{1} = 1$ and $\chi_{2} = 3$ and $\chi_{1} = 1$ and $\chi_{2} = 4/3$ respectively.

These features can be explained on a case by case basis as before by examining the periodicity properties of the unitary time evolution operator at appropriate instants. It is however evident that the simple inference that an $l$-subpacket fractional revival is associated with an $l$-strand tomogram alone does not hold when more than one time scale is involved in the Hamiltonian, and there can be several tomograms possible at a given instant depending on the interplay between the different time scales in the system.

\subsection{Squeezing and higher-order squeezing}
We now proceed to examine the squeezing and higher-order squeezing properties of the system with Hamiltonian $H$. Once again, our aim is to identify and assess  these nonclassical effects directly from the tomogram without attempting to reconstruct the state of the system at any instant of time.  The extent of quadrature squeezing at a given instant is essentially determined by the numerical value of the variance of the quadrature observables. The state of the system is said to be squeezed in $x$  if the variance 
$(\Delta x)^{2}=\langle (x- \langle {x} \rangle)^{2} \rangle$ is less than the variance  of $x$ in a CS  $\ket{\alpha}$.  Generalization of this definition to include higher-order squeezing allows for two possibilities.  Hong-Mandel squeezing of order $2 q$ in $x$ requires that  $\langle (x- \langle {x} \rangle)^{2q} \rangle$
  for the given state  is less than the corresponding expectation value for the CS.  Here $q = 2,3,\dots$. 
  We will estimate the extent of Hong-Mandel squeezing  by calculating the central moments of the probability distribution corresponding to the appropriate  quadrature. For example, if we wish to determine  the extent of second-order Hong-Mandel squeezing in the $x$ quadrature, we simply calculate the fourth central moment of a horizontal cut of the tomogram
 at $\theta=0$. Note that for $q=1$, Hong-Mandel squeezing is identical to quadrature squeezing. 

  Hillery type squeezing of order $q$ corresponds  to squeezing in either of the pair of operators 
  $Z_{1} =( a^{q} + {a^{\dagger}}^{q})/ \sqrt{2}$ and  $Z_{2} =( a^{q} - {a^{\dagger}}^{q})/ \sqrt{2} \, \rmi$, 
  where $q = 2,3,\dots$.  A useful quantifier $D_{q}$ of $q$th-power squeezing in $Z_{1}$ for instance, is defined  \cite{Hillery} in terms of the commutator 
 $[a^{q}, {a^{\dagger}}^{q}] = F_{q}(N)$  as 
\begin{equation}
D_{q}= \frac{\aver{(\Delta Z_{1})^{2}} - |F_{q}(N)|}{|{F_{q}(N)}|}.
\label{eqn:squeeze_param}
\end{equation}
where $(\Delta Z_{1})^{2}$ is the variance in $Z_{1}$. A similar definition holds for $q$th power squeezing in $Z_{2}$.  
We note that $F_{q}(N)$ is a polynomial function of  order $(q - 1)$ in $N$. A state is $q$th-power squeezed if $-1 \leq D_{q} < 0$.
 It is clear that  $Z_1$ and $Z_2$ cannot be obtained in a straightforward manner from the tomogram as they involve terms with products of powers of different rotated  quadratures and hence cannot be assigned probability distributions directly from a set of tomograms.

 However an illustrative treatment of the problem of expressing the expectation value of  a product of moments of  creation and destruction operators in terms of the tomogram $w$ and Hermite polynomials  \cite{wunsche} leads to the result
\begin{eqnarray}
	\label{eqn:wunsche}
\nonumber \braket{a^{\dagger k} a^l} &= C_{k l} \sum_{m=0}^{k+l} \exp \left(-\rmi(k-l)\left(\frac{m\pi}{k+l+1}\right)\right) \\
& \int_{-\infty}^{\infty} \rmd X_\theta  \ w\left(X_\theta, \ \frac{m\pi}{k+l+1}\right) H_{k+l}\left( X_{\theta}  \right) ,
\end{eqnarray}
where
\begin{equation}
\nonumber C_{k l}=\frac{k! l!}{(k+l+1)! \sqrt{2^{k+l}}}.
\end{equation}

This form is readily amenable to numerical computations.
We therefore need to consider $(k+l+1)$ values of the tomogram variable $\theta$  in order to calculate a moment of order $(k+l)$.  In a single tomogram, this amounts to using  $(k+l+1)$  probability distributions 
$w(X_{\theta}$) corresponding to  these chosen values of $\theta$,  in order to  calculate the squeezing parameter $D_{q}$  from the tomogram without resorting to detailed state reconstruction.  As the system evolves in time, the extent of squeezing at various instants are determined from the instantaneous tomograms in this manner.

\begin{figure}
	\centering    
	\includegraphics[width=0.48\textwidth]{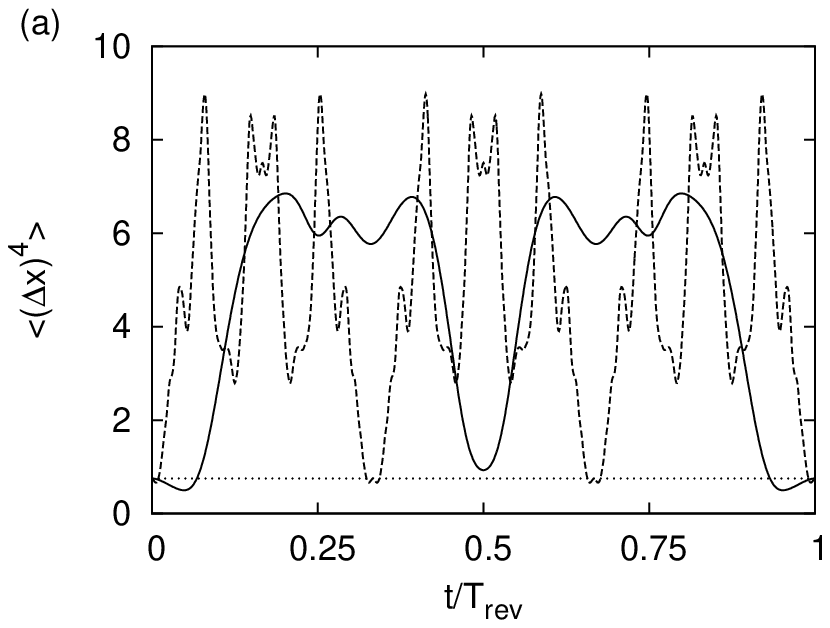}
	\includegraphics[width=0.48\textwidth]{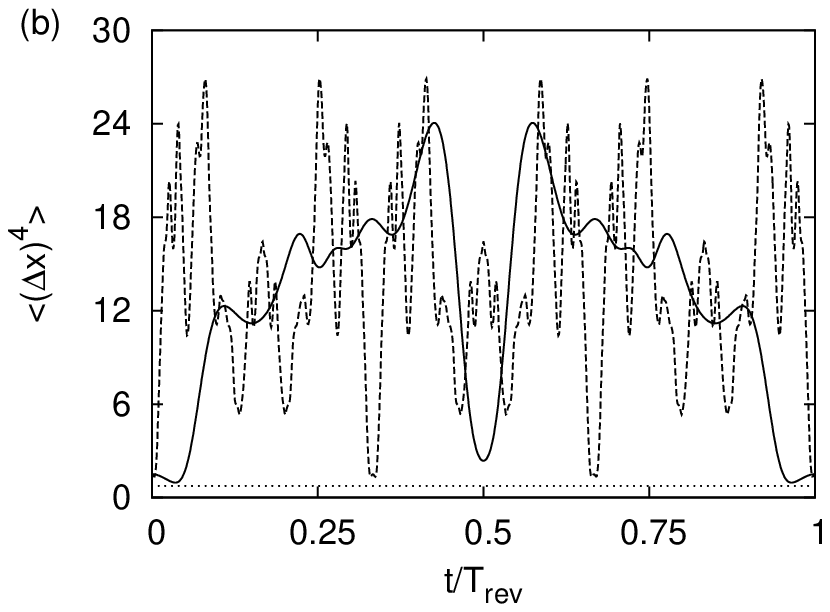}
	\caption{Hong-Mandel squeezing  as a function of scaled time $t/T_{rev}$ for initial  states (a)$\ket{\alpha}$ and (b)$\ket{\alpha,1}$ with $\alpha = 1$.  The solid (dashed) line corresponds to the Kerr (respectively cubic) Hamiltonian.}
	\label{fig:squeezingcspacshongmandel}
\end{figure}

For the squeezed vacuum, $\ket{\alpha}$ and $\ket{\alpha,1}$  we have verified that the variance and hence the Hong-Mandel (equivalently Hillery type squeezing) properties inferred directly from tomograms are in excellent agreement with  corresponding results obtained analytically by calculating the variance from the state. 
We have also computed $\aver{(\Delta x)^{4}}$ (equivalently the  higher-order Hong-Mandel squeezing parameter)  directly from the tomogram for initial states $\ket{\alpha}$ and $\ket{\alpha,1}$ evolving under the Kerr-like Hamiltonian and the  cubic Hamiltonian ${a^{\dagger}}^{3}a^{3}$,  setting $\alpha = 1$ in both cases  without loss of generality (see figures~\ref{fig:squeezingcspacshongmandel}(a) and (b)).
From these figures it is evident that independent of the precise nature of the initial field state $\aver{(\Delta x)^{4}}$ oscillates more rapidly in the case of the cubic Hamiltonian compared to the Kerr-like Hamiltonian. Earlier these squeezing properties  have  been investigated  for an initial CS and PACS evolving under a Kerr-like Hamiltonian  by calculating $\aver{(\Delta x)^{4}}$ explicitly for the state at different times \cite{sudhsqueezing}. Our results from the tomograms for the Kerr-like Hamiltonian are in excellent agreement with these.
 
\begin{figure*}
	\centering
	\includegraphics[width=0.3\textwidth]{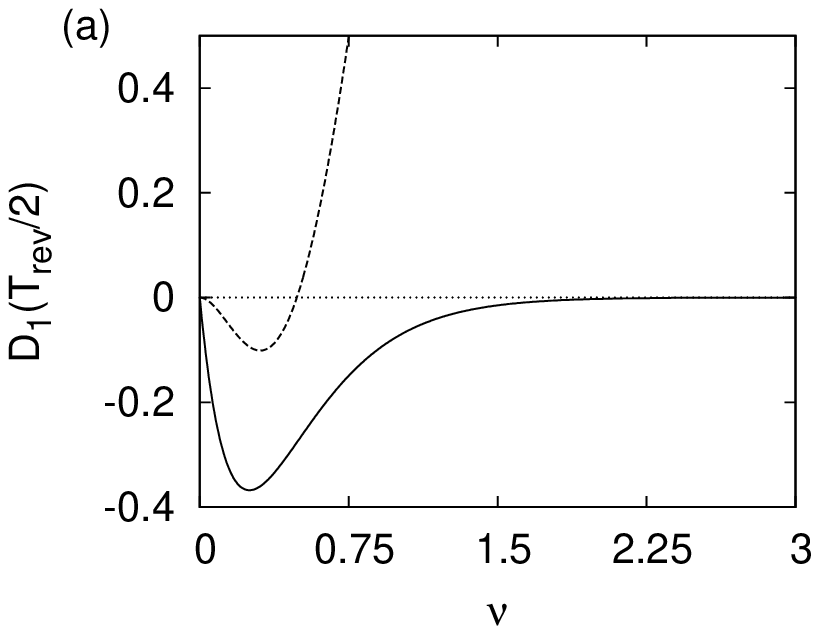}
	\includegraphics[width=0.3\textwidth]{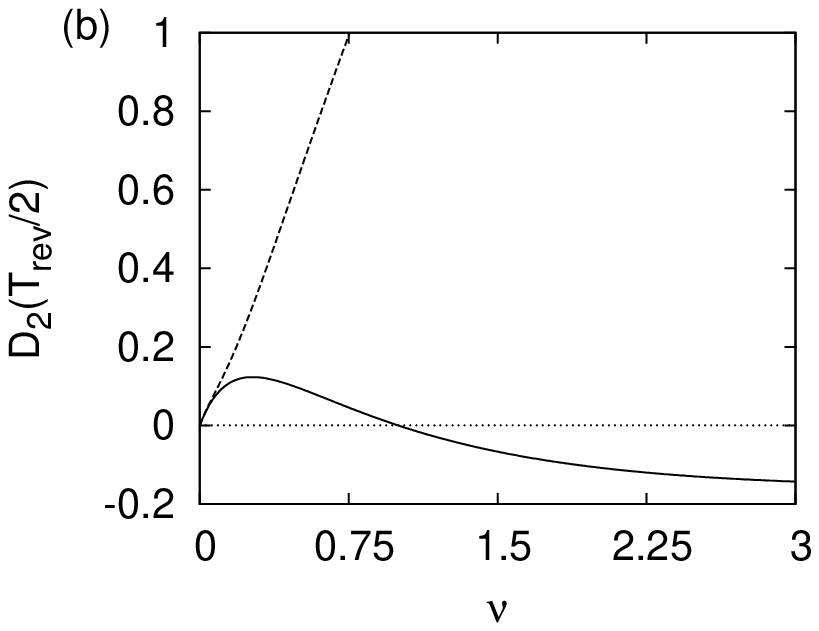}
	\includegraphics[width=0.3\textwidth]{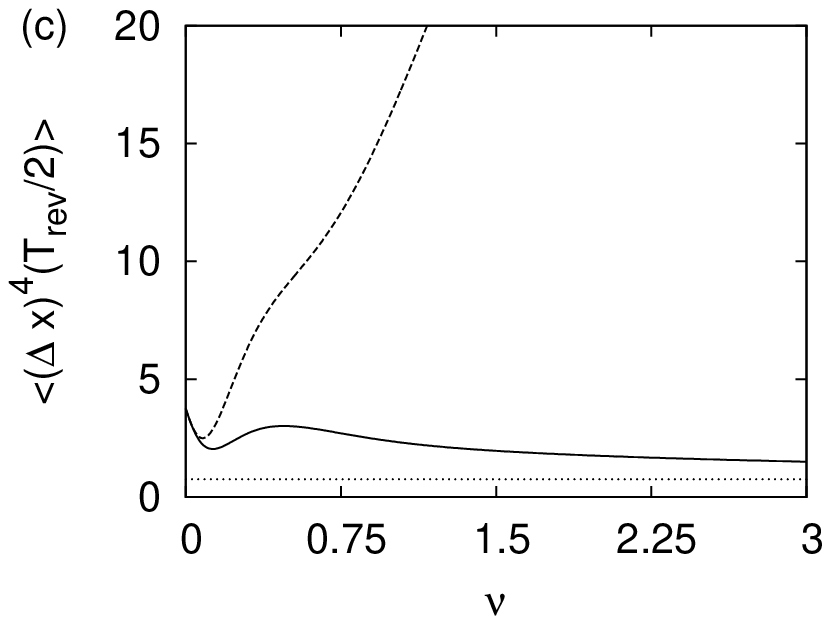}
	\caption{ (a) $D_{1}$ for initial $\ket{\alpha}$, (b) $D_{2}$ for initial $\ket{\alpha,1}$, and (c) $\langle (\Delta x)^{4} \rangle$ for initial $\ket{\alpha,1}$ versus $\nu$ for real $\alpha$ at instant $\frac{T_{rev}}{2}$. The solid (dashed) line corresponds to the Kerr (respectively cubic) Hamiltonian.}
	\label{fig:squeezingcspacshillery}
\end{figure*}

We have also examined the manner in which the squeezing properties depend on the numerical value of $\alpha$, directly from the relevant tomograms. For real $\alpha$ in the range $0$ to $\sqrt{3}$ we have plotted $D_{1}$  the quantifier of the extent of squeezing  versus $\nu = |\alpha|^{2}$ at the instant $T_{rev}/2$ for an initial CS evolving under both the Kerr and cubic Hamiltonians
(\fref{fig:squeezingcspacshillery}(a)). It is evident that the state is squeezed for a larger range of  values of $\nu$ in the case of the Kerr-like  Hamiltonian than for the cubic Hamiltonian.  Further the extent of squeezing as measured by the numerical value of $D_{1}$ is more for a given $\nu$ in the former case as compared to the latter. 

For an initial state $\ket{\alpha,1}$ we have computed the extent of higher-order Hillery type squeezing from the tomograms over the range $0 \leq \nu \leq 3$, at $T_{rev}/2$ for both Hamiltonians. While $D_{2}$ is not negative for any $\nu$ in this range for the cubic Hamiltonian, it becomes negative for $\nu \geq 0.8$ approximately for the Kerr-like Hamiltonian (\fref{fig:squeezingcspacshillery}(b)). In contrast $\aver{(\Delta x)^{4}}$  is not negative  in both cases over this range of values of $\nu$ (\fref{fig:squeezingcspacshillery}(c)).

\section{The double-well BEC: A tomographic approach \label{Sec3}}
 \subsection{The Revival Phenomena}
We now proceed to examine nonclassical effects in a BEC condensed in a double-well potential. The effective Hamiltonian of this bipartite system is ~\cite{sanz}
\begin{equation}
H_{bec} =\omega_{0} N_{tot} + \omega_{1} (a^{\dagger}a - b^{\dagger}b)
+ U_{ab} N_{tot}^{2} - \lambda (a^{\dagger} b + a b^{\dagger}).
\label{eqn:2_mode_Hamiltonian}
\end{equation}
$(a, a^{\dagger})$ and $(b, b^{\dagger})$ are the boson annihilation and creation operators corresponding to the two subsystems $A$ and $B$ which comprise the atomic species condensed in the two wells. They 
satisfy $[a,a^{\dagger}] = 1$, $[b,b^{\dagger}] = 1$ with all other commutators vanishing.  Here, $N_{tot}=(a^{\dagger}a+b^{\dagger}b)$ and $U_{ab}$ is the strength of the nonlinearity.  It is convenient to define 
the  effective  interaction strength  by  the parameter $\lambda_{1}=\sqrt{\omega_{1}^{2}+\lambda^{2}}$. 
It is easy to see that  $[H_{bec},N_{tot}]=0$.

Before  we examine revivals and squeezing phenomena in this system by investigating relevant tomograms,
we recall that we now have two quadrature operators to consider, one for each subsystem. The tomogram is therefore denoted by $w(X_{\theta_{1}}, \theta_{1}; X_{\theta_{2}}, \theta_{2})$ with subscripts $1$ and $2$ corresponding to subsystems $A$ and $B$ respectively. 

Analogous to the single-mode example  we now consider a pure state $\ket{\psi}$  expanded in the Fock bases $\{\ket{m}\}$, $\{\ket{n}\}$  corresponding to subsystems A and B respectively as  
$\ket{\psi}  = \sum_{m,n=0}^{\infty} c_{mn} \ket{m; n}$ , where $\ket{m;n}$ is a short-hand notation 
for $(\ket{m} \otimes \ket{n})$ and $c_{mn}$ are the expansion coefficients.

It is straightforward to extend the procedure for expressing the optical tomogram in the single-mode example in terms of Hermite polynomials to a generic bipartite system whose subsystems are infinite-dimensional.  We can see that  analogous to  \eref{eqn:tomogfock}, in this case we have
\begin{eqnarray}
\nonumber w(X_{\theta_{1}} , \theta_{1}; X_{\theta_{2}}, \theta_{2})
= \frac{\exp(-X_{\theta_{1}}^{2}-X_{\theta_{2}} ^{2})}{\pi}\\
 \times\left| \sum_{m,n=0}^{\infty} \frac{c_{mn} \rme^{- \rmi (m \theta_{1} + n \theta_{2})}}{\sqrt{m! n! 2^{m+n}}} H_{m} (X_{\theta_{1}}) H_{n}(X_{\theta_{2}}) \right|^{2}.
\label{eqn:2_mode_opt_tomo}
\end{eqnarray}
The tomograms corresponding to the two subsystems (reduced tomograms) are given  by 
\begin{equation}
w_{1}(X_{\theta_{1}}, \theta_{1})= \int_{-\infty}^{\infty} w(X_{\theta_{1}}, \theta_{1}; X_{\theta_{2}}, \theta_{2}) \rmd X_{\theta_{2}},
\label{eqn:tomo_2_mode_sub_A}
\end{equation}
for any fixed value of $\theta_{2}$ and
\begin{equation}
w_{2}(X_{\theta_{2}}, \theta_{2})= \int_{-\infty}^{\infty} w(X_{\theta_{1}}, \theta_{1}; X_{\theta_{2}}, \theta_{2}) \rmd X_{\theta_{1}}
\label{eqn:tomo_2_mode_sub_B}
\end{equation}
for any fixed value of $\theta_{1}$.

We denote by $\ket{\alpha_{a}}$ (respectively $\ket{\alpha_{b}}$) the CS  formed from the condensate corresponding to subsystem A (respectively B) 
and by $\ket{\alpha_{a},1}$ (respectively $\ket{\alpha_{b},1}$) a 1-boson added CS  corresponding to subsystem A (respectively B). 
The initial states considered  by us are  factored product states  of the form 
  $\ket{\alpha_{a}} \otimes \ket{\alpha_{b}}$ (denoted by $\ket{\psi_{00}}$), 
  $\ket{\alpha_{a},1} \otimes \ket{\alpha_{b},1}$( denoted by $\ket{\psi_{11}}$) and
 $ \ket{\alpha_{a},1} \otimes \ket{\alpha_{b}}$ (denoted by $\ket{\psi_{10}}$).

The state at a subsequent time $t$ is  entangled in general  and  corresponding to the 
initial state $\ket{\psi_{00}}$ we have \cite{sanz}
\begin{eqnarray}
\nonumber \ket{\psi_{00} (t)} &= \exp (-n/2) \sum_{p,q=0}^{\infty} \frac{(\alpha(t))^{p} (\beta(t))^{q}}{\sqrt{p! q!}}\\
& \exp (- \rmi t (\omega_{0} (p+q)+ U_{ab} (p + q)^{2})) \ket{p;q}.
\label{eqn:Psi_t}
\end{eqnarray}
Here, 
\begin{equation}
\nonumber \alpha (t) = \alpha_{a} \cos (\lambda_{1} t) + \rmi \frac{\sin (\lambda_{1} t)}{\lambda_{1}} \left( \lambda \alpha_{b} - \omega_{1} \alpha_{a} \right),\\
\label{eqn:alpha_t}
\end{equation}
\begin{equation}
\nonumber \beta (t) = \alpha_{b} \cos (\lambda_{1} t) + \rmi \frac{\sin (\lambda_{1} t)}{\lambda_{1}} \left( \lambda \alpha_{a} + \omega_{1} \alpha_{b} \right),
\label{eqn:beta_t}
\end{equation}
and $n =  ({|\alpha_{a}|}^{2} + {|\alpha_{b}|}^{2})$ \cite{sanz}.

We can use a similar  procedure to obtain $ \ket{\psi_{10}(t)}$ and  $ \ket{\psi_{11}(t)}$.
 We have
\begin{eqnarray}
\nonumber \ket{\psi_{10} (t)} = \frac{1}{d_{10}}  &\left(a^{\dagger} \lambda_{1} \cos (\lambda_{1} t) +  \rmi (\lambda b^{\dagger} - \omega_{1} a^{\dagger}) \sin (\lambda_{1} t) \right)\\
& \times \exp(- \rmi U_{ab} (2 N + 1) t) \ket{\psi_{00} (t)},
\label{eqn:Psi_10_t}
\end{eqnarray}
and
\begin{eqnarray}
\nonumber \ket{\psi_{11} (t)} = &\frac{1}{d_{11}} \biggl( 2 \omega_{1}^{2} \, a^{\dagger} b^{\dagger} +\omega_{1} \lambda \left( a^{\dagger 2}- b^{\dagger 2} \right) \\
\nonumber &+ \cos (2 \lambda_{1} t) \left(2 \lambda^{2} \, a^{\dagger} b^{\dagger} - \omega_{1} \lambda \left( a^{\dagger 2}-b^{\dagger 2} \right) \right)\\
\nonumber &+ \rmi \sin (2 \lambda_{1} t) \lambda \lambda_{1} (a^{\dagger 2} +  b^{\dagger 2}) \biggr)\\
& \times \exp (- 4 \rmi U_{ab} ( N + 1) t) \ket{\psi_{00} (t)}.
\label{eqn:Psi_11_t}
\end{eqnarray}
Here $d_{10}=  \lambda_{1} \exp (\rmi \omega_{0} t) \sqrt{1+{|\alpha_{a}|}^{2}}$ and $d_{11} = 2 \lambda_{1}^{2} \exp (2 \rmi \omega_{0} t) \sqrt{1+ {|\alpha_{a}|}^{2}}\sqrt{1+{|\alpha_{b}|}^{2}}$. 

We have obtained a form for the density matrix which is very useful in numerical computations. The salient steps in the calculation of the density matrix $\rho_{m_{1},m_{2}}$ corresponding to the factored product state 
 $\ket{\alpha_{a},m_{1}} \otimes \ket{\alpha_{b},m_{2}}$ for any positive integer value of $m_{1}$ and $m_{2}$ are outlined in the Appendix.  We have shown there that  $\rho_{m_{1},m_{2}}$ can be expressed in terms of an operator $M_{m_{1},m_{2}}(t)$ and $ \ket{\psi_{00}(t)}$ as
\begin{equation}
\rho_{m_{1},m_{2}} (t) = M_{m_{1},m_{2}}(t) \ket{\psi_{00}(t)}\bra{\psi_{00}(t)} M^{\dagger}_{m_{1},m_{2}}(t).
\label{rho_numerics}
\end{equation}
Defining $p_{max}=(k+m_{2}-l)$ and $q_{max}=(l+m_{1}-k)$ we have

\begin{eqnarray}
\nonumber &M_{m_{1},m_{2}}(t)= \frac{1}{\kappa} \biggl[\sum_{k=0}^{m_{1}}\sum_{l=0}^{m_{2}}\sum_{p=0}^{p_{max}}\sum_{q=0}^{q_{max}}(-1)^{k-p} {m_{1} \choose k} {m_{2} \choose l} \\
\nonumber &\hspace{2 em} {p_{max} \choose p} {q_{max} \choose q} \exp(-\rmi\lambda_{1} t (2(k-l)+m_{2}-m_{1}))\\ 
\nonumber &\hspace{2 em} (\cos(\gamma/2))^{(k+l+p+q)} (\sin(\gamma/2))^{(2(m_{1}+m_{2})-(k+l+p+q))}\\
\nonumber &\hspace{2 em} a^{\dagger (p+q_{max}-q)} b^{\dagger (q+p_{max}-p)} \biggr] \exp(-\rmi\omega_{0} t(m_{1}+m_{2}))\\
& \times \exp(-\rmi U_{ab} t (m_{1}+m_{2}) (2 N_{tot} + m_{1}+m_{2})).
\label{eqn:intermed_rho_numerics}
\end{eqnarray}
Here $\kappa=\sqrt{m_{1}! L_{m_{1}}(-{|\alpha_{a}|}^{2}) m_{2}! L_{m_{2}}(-{|\alpha_{b}|}^{2})}$ and  $\gamma=\cos^{-1}(\omega_{1}/\lambda_{1})$. $L_{m}$ are the Laguerre polynomials which appear in the normalization of $\ket{\alpha,m}$. This expression for the  general density matrix can be easily seen to reduce to the forms needed in our case where the states are $\ket{\psi_{00}(t)}$, $\ket{\psi_{11}(t)}$ and  $\ket{\psi_{10}(t)}$.

We now proceed to examine the revival phenomena by studying appropriate tomograms in this case.
A straightforward calculation reveals that full and fractional revivals occur provided $\omega_{0}=m U_{ab}$, $\lambda_{1}=m' U_{ab}$, $m,m' \in \mathbb{Z}$, and $(m+m')$ is odd. Fractional revivals occur at fractions of the revival period $T_{rev}$ (which is equal to $\pi /U_{ab}$).  This follows (similar to the single-mode case) from the periodicity property of $\exp(-i U_{ab} N_{tot}^{2})$. For instance, we can easily show that an initial state $\ket{\psi_{00}}$ evolves at an instant $\pi/(s U_{ab})$  ($s$: even integer) to
\begin{eqnarray}
\nonumber \ket{\psi_{00} (\pi / s U_{ab})} = &\sum_{j=0}^{s-1}  a_{j} \ket{\alpha(\pi/s U_{ab}) \rme^{- \rmi \pi(m + 2 j)/s}} \\
& \otimes \ket{\beta(\pi/s U_{ab}) \rme^{- \rmi \pi (m + 2 j)/s}},
\label{eqn:frac_rev_even}
\end{eqnarray}
If $s$ is an odd integer,
\begin{eqnarray}
\nonumber \ket{\psi_{00} (\pi/s U_{ab})} = &\sum_{j=0}^{s-1} b_{j} \ket{\alpha(\pi/s U_{ab}) \rme^{- \rmi \pi(m + 2 j + 1)/s}} \\
& \otimes \ket{\beta(\pi/s U_{ab}) \rme^{- \rmi \pi(m + 2 j + 1) / s}}.
\label{eqn:frac_rev_odd}
\end{eqnarray}
Note that $\alpha\left( \pi/(s U_{ab}) \right)$ and $\beta\left( \pi/(s U_{ab}) \right)$ are obtained from \eref{eqn:alpha_t} and  \eref{eqn:beta_t}.

 We recall that in the single-mode system the occurrence of full and fractional revivals are related to the presence of distinct strands in the tomogram. In this bipartite system however, the tomogram is a 4-dimensional hypersurface. Hence we need to consider appropriate sections to identify and examine nonclassical effects. The 2-dimensional section ($X_{\theta_{2}}$ - $X_{\theta_{1}}$) obtained by setting $\theta_{1}$ and $\theta_{2}$ constant is a natural choice for investigating  not only revival phenomena but also squeezing properties. In contrast to the single-mode case where strands appear in the tomograms these sections are characterized by  `blobs'  at instants of fractional revivals. The number of blobs gives the number of subpackets in the wave packet. 
 
\begin{figure}
\centering
\includegraphics[scale=0.7]{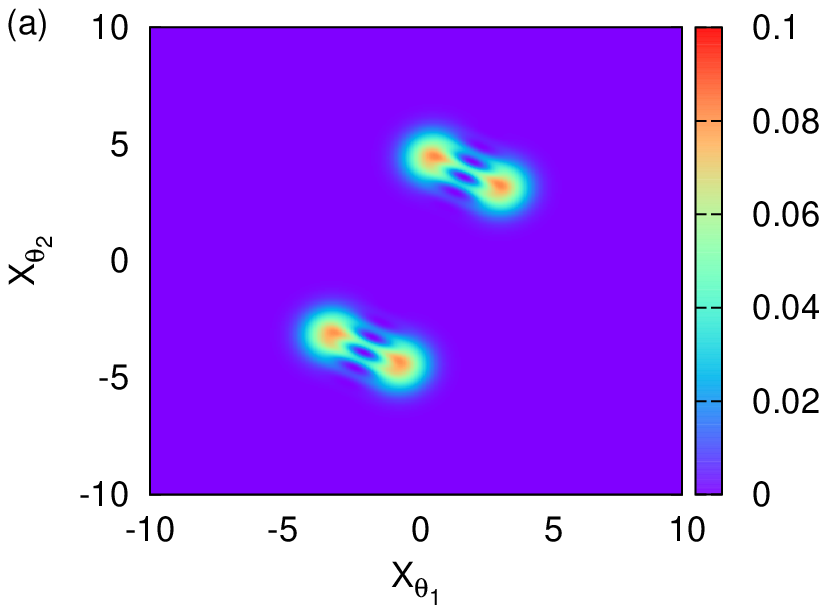}
\includegraphics[scale=0.7]{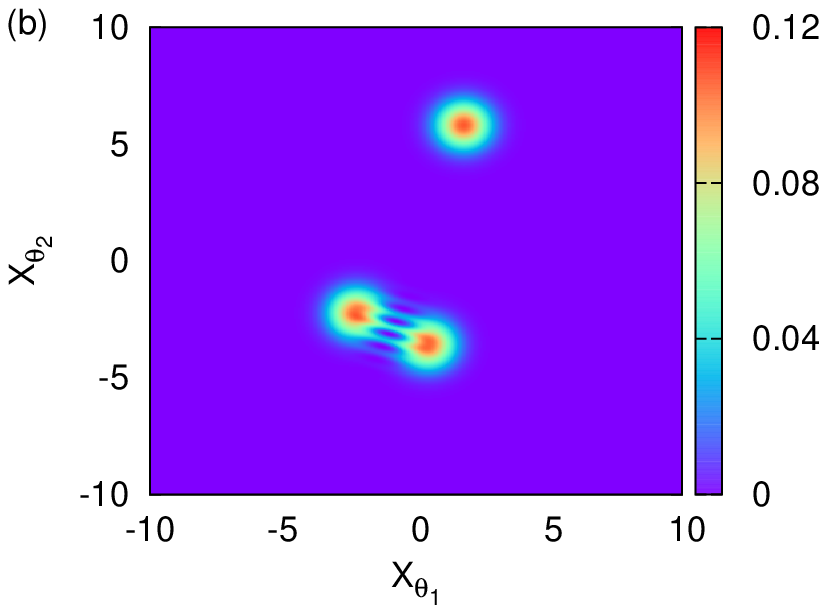}
\includegraphics[scale=0.7]{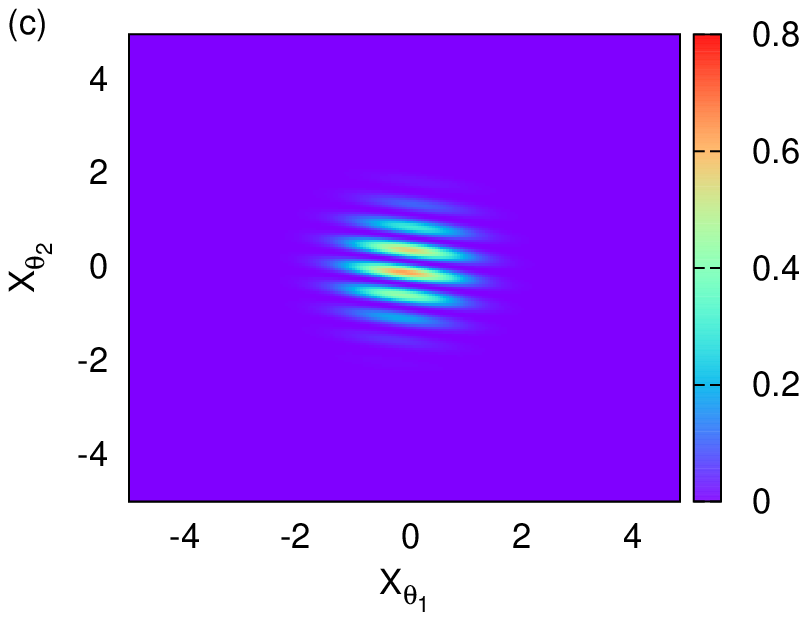}
\includegraphics[scale=0.7]{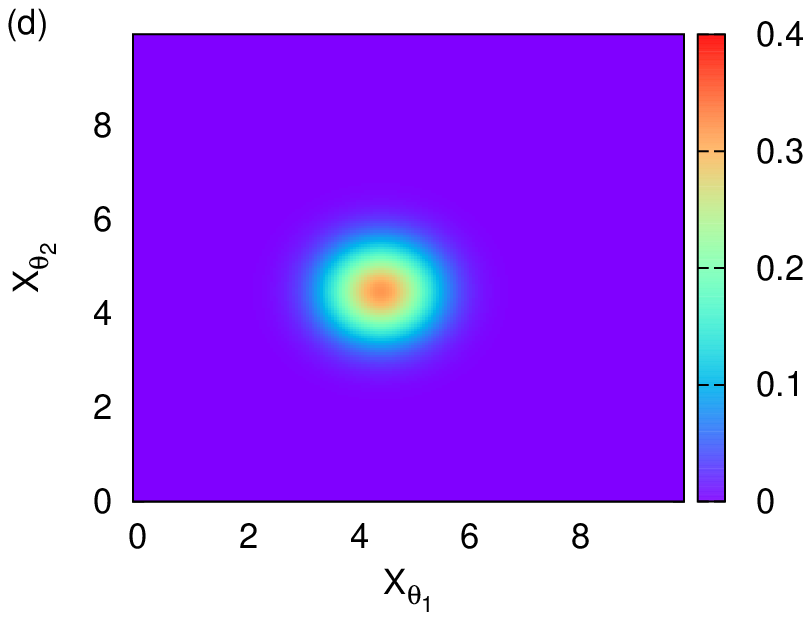}
\caption{Sections of the optical tomogram for $\theta_{1}=\theta_{2}=0$ at instants of fractional 
revivals (a)$T_{rev}/4$, (b)$T_{rev}/3$, (c)$T_{rev}/2$, and (d)$T_{rev}$. $\alpha_{a}=\alpha_{b}= \sqrt{10}$, for an initial state $\ket{\psi_{00}}$.}
\label{fig:2_mode_frac_rev}
\end{figure}

 Although $\ket{\alpha}$ is expanded as an infinite superposition of photon number states, in practice, a numerical computation can  be carried  out using only a large but finite sum of these basis states. An alternative is to use truncated coherent state(TCS)  \cite{tcs} instead of the standard CS. The latter are  defined as 
\begin{equation}
{\ket{\alpha}}_{tcs} = \frac{1}{\sum_{n=0}^{N_{max}}( |\alpha|^{2 n}/n!)} \sum_{p=0}^{N_{max}} \frac{\alpha^{p}}{\sqrt{p!}} \ket{p}.
\label{eqn:TCS}
\end{equation}
 where $N_{max}$ is a sufficiently large but finite integer. We have worked with an
  initial state $\ket{\alpha_{a}}_{tcs} \otimes \ket{\alpha_{b}}_{tcs}$  instead of $\ket{\psi_{00}}$ and verified that 
as  this state  evolves under $H_{bec}$  the revival phenomena and squeezing properties mimic that of $\ket{\psi_{00}}$ remarkably well.  Hence we have used initial  states $\ket{\psi_{00}}$,  $\ket{\psi_{10}}$  and $\ket{\psi_{11}}$  for our numerical computations. We set $\theta_{1} = \theta_{2} = 0$, $\omega_{0}=10$, $\omega_{1}=3$, $\lambda=4$, $U_{ab}=1$, and $\alpha_{a} = \alpha_{b} = 10$ in figures \ref{fig:2_mode_frac_rev}  (a)-(d) where we present tomograms corresponding to different  fractional revivals for the initial state $\ket{\psi_{00}}$.  At instants $T_{rev}/4$,  $T_{rev}/3$ and $T_{rev}$  (figures \ref{fig:2_mode_frac_rev}(a), (b) and (d) respectively) we see 4, 3 and a single blob in the tomogram  along with interference patterns as expected for this choice of values of $\lambda_{1}$ and $\omega_{0}$ for they satisfy  the conditions $\omega_{0}=m U_{ab}$, $\lambda_{1}=m' U_{ab}$, $m,m' \in \mathbb{Z}$, and $(m+m')$ is odd, necessary for the revival phenomena to occur. 
However in \fref{fig:2_mode_frac_rev}(c)  corresponding to the instant $T_{rev}/2$  blobs are absent  and we  merely see interference patterns.
 This is primarily due to the specific choice of  {\it  real} values of  $\alpha_{a}$ and $\alpha_{b}$  as explained below. At the instant $T_{rev}/2$ it follows from \eref{eqn:Psi_t} that the state of the system can be expanded in terms of superpositions of factored products of CS  corresponding to $A$ and $B$ as 
\begin{eqnarray}
\nonumber \ket{\psi_{00}(T_{rev}/2)}= &\frac{(1-\rmi)}{2} \ket{{- 2 \rmi}/{\sqrt{10}}} \otimes \ket{{- 14 \rmi}/{\sqrt{10}}} \\
&+ \frac{(1+\rmi)}{2} \ket{{2 \rmi}/{\sqrt{10}}} \otimes \ket{{14 \rmi}/{\sqrt{10}}}.
\label{eqn:Psi_half_rev_sp_param}
\end{eqnarray}
It is  now straightforward to see why the interference patterns alone appear in the tomogram, as a simple calculation gives 
\begin{eqnarray}
\nonumber |\psi_{00}(X_{10},X_{20})|^{2} = |\langle X_{10},0;X_{20},0 | \psi_{00}(T_{rev}/2)\rangle|^{2}\\
=\frac{1}{\pi} \rme^{-(X_{10}^{2} + X_{20}^{2})} \left(1-\sin(4 (X_{10} + 7 X_{20})/ \sqrt{5} )\right). 
\label{eqn:2_mode_Trevby2_interference}
\end{eqnarray}
Here  $X_{10}$ denotes $X_{\theta_{1}}$ for $\theta_{1}=0$ and $X_{20}$  denotes $X_{\theta_{2}}$ for $\theta_{2}=0$.
In contrast, it can be seen that if $\alpha_{a}$ and $\alpha_{b}$  were chosen to be  generic complex numbers two blobs together with the interference pattern would be seen to appear at $T_{rev}/2$ also.   

Similar results hold in the case of initial states $\ket{\psi_{11}}$ and $\ket{\psi_{10}}$.

\subsection{Squeezing and higher-order squeezing of the condensate}

\begin{figure}
\centering
\includegraphics[width=0.49\textwidth]{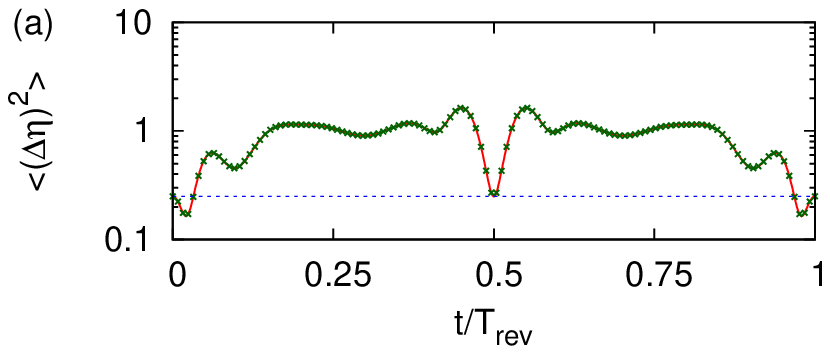}
\includegraphics[width=0.49\textwidth]{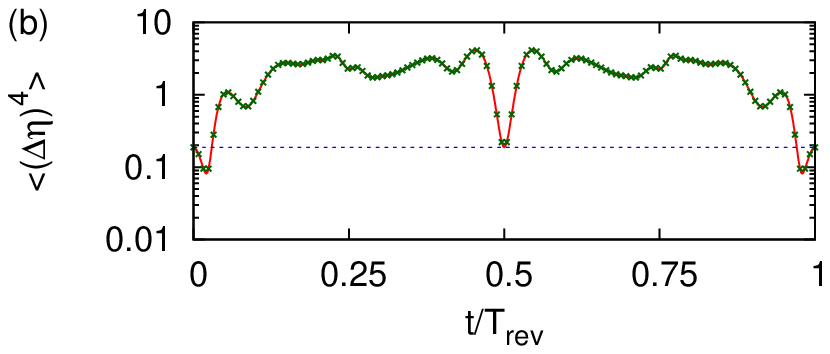}
\includegraphics[width=0.49\textwidth]{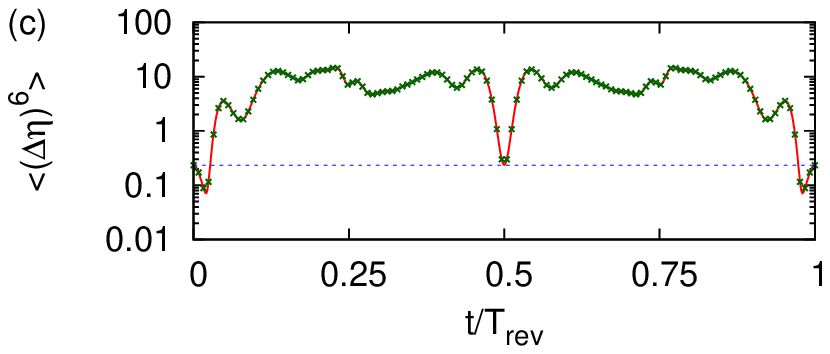}
\includegraphics[width=0.49\textwidth]{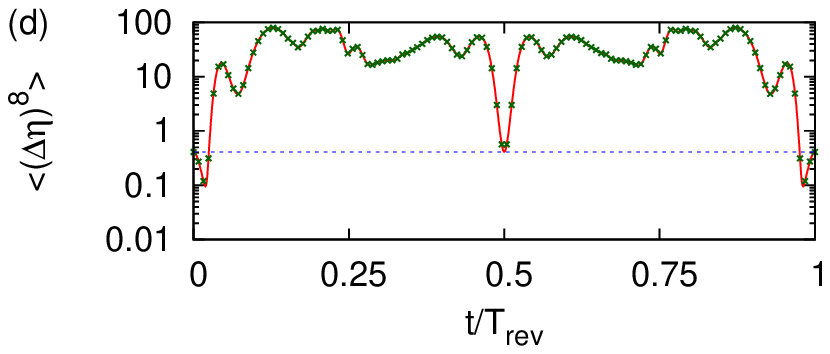}
\caption{$2 q$-order moment of $\eta=(a+a^{\dagger}+b+b^{\dagger})/2\sqrt{2}$ 
versus  $t/T_{rev}$ with $|\alpha_{a}|= |\alpha_{b}|=1$, for (a) $q=1$, (b) $q=2$, (c) $q=3$, and (d) $q=4$. Moments obtained directly from the tomogram are denoted by green crosses and those from the state by the red solid curve. The horizontal line denotes the moment corresponding to the CS.}\label{fig:2_mode_hong_time_dep}
\end{figure}

The extent of Hong-Mandel squeezing is simply obtained  as in the single-mode example,  by calculating the central moments of the probability distribution corresponding to the quadrature. We examine two-mode squeezing  by evaluating appropriate moments of the quadrature variable $\eta=(a+a^{\dagger}+b+b^{\dagger})/2\sqrt{2}$.  These are obtained from the $\theta_{1}=\theta_{2}=0$ section of the tomogram as the system evolves in time. These moments have also been obtained by explicit calculation of the relevant expectation values of $\eta$ in the state of the system at different times. In both cases the initial state considered is $\ket{\psi_{00}}$.
In figures~\ref{fig:2_mode_hong_time_dep} (a)-(d), the variance and $2 q$-order moments for $q = 1,2,3$ and $4$ obtained  both from the states and directly from the tomogram are plotted as a function of time. It is evident that they are in excellent agreement with each other at all instants. The horizontal  line in each figure denotes the value below which the state is squeezed. For all values of $q$ considered,  the state is squeezed (higher-order squeezed)  in the neighborhood of revivals, and the actual magnitude at various instants is considerably sensitive to the value of $q$ as expected. We have also verified that the extent of this Hong-Mandel squeezing depends on the magnitude of $\alpha_{a}$ and $\alpha_{b}$ at various instants of time.

For numerical computation  of the Hillery type higher-order squeezing recall that in the single-mode case we  used the expression \cite{wunsche} for moments of the creation and destruction operators  in terms of the tomogram and Hermite polynomials given by 
\begin{eqnarray}
	\label{eqn:Wunsche}
\nonumber \braket{a^{\dagger k} a^l} &= C_{k l} \sum_{m=0}^{k+l} \exp \left(-\rmi(k-l)\left( \frac{m\pi}{k+l+1}\right)\right) \\
& \int_{-\infty}^{\infty} \rmd X_\theta  \ w\left(X_\theta, \frac{m\pi}{k+l+1}\right) H_{k+l}\left( X_{\theta}  \right) ,
\end{eqnarray}
where
\begin{equation}
\nonumber C_{k l}=\frac{k! l!}{(k+l+1)! \sqrt{2^{k+l}}}.
\end{equation}

A straightforward extension to the two-mode system gives us the  required expression
\begin{eqnarray}
\nonumber \langle a^{\dagger k} a^{l} &b^{\dagger m} b^{n} \rangle =  c_{k l m n} \sum_{p=0}^{k+l} \sum_{q=0}^{m+n} \exp \left(-\rmi (k-l)\theta_{1 p} \right)\\
\nonumber & \exp\left( - \rmi (m-n)\theta_{2 q}\right) \int_{-\infty}^{+\infty} \rmd X_{\theta_{1 p}} \int_{-\infty}^{+\infty} \rmd X_{\theta_{2 q}} \\
&\hspace*{-1.5 em}w\left(X_{\theta_{1 p}}, \theta_{1 p}; X_{\theta_{2 q}}, \theta_{2 q}\right) H_{k+l}(X_{\theta_{1 p}}) H_{m+n}(X_{\theta_{2 q}}).
\label{eqn:2_mode_Wunsche}
\end{eqnarray}
Here
$$c_{k l m n}=\frac{k! l! m! n!}{(k+l+1)! (m+n+1)! \sqrt{2^{k+l+m+n}}},$$ $\theta_{1 p}= \frac{p \pi}{k+l+1}$, 
and $\theta_{2 q} = \frac{q \pi}{m+n+1}$. Note that $(k+l+1)(m+n+1)$ gives the number of 2- dimensional slices of the tomogram that are required to calculate $\langle a^{\dagger k} a^{l} b^{\dagger m} b^{n} \rangle$.

\begin{figure}
\centering
\includegraphics[width=0.49\textwidth]{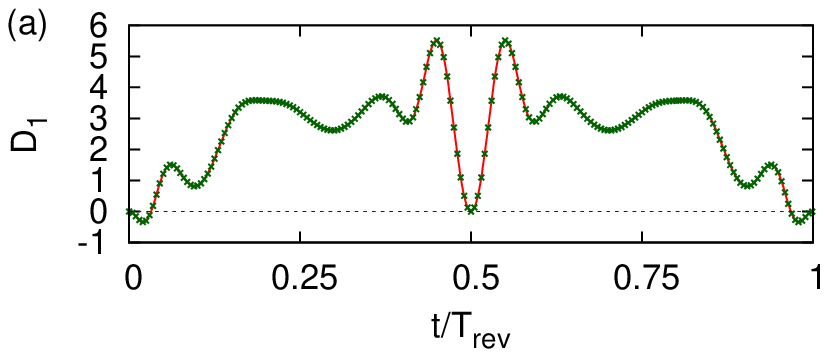}
\includegraphics[width=0.49\textwidth]{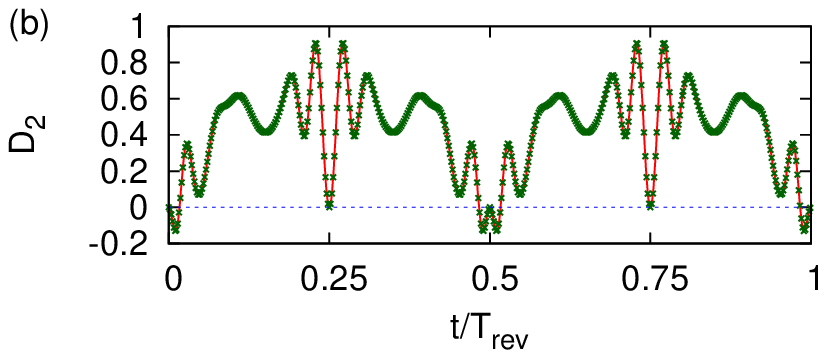}
\includegraphics[width=0.49\textwidth]{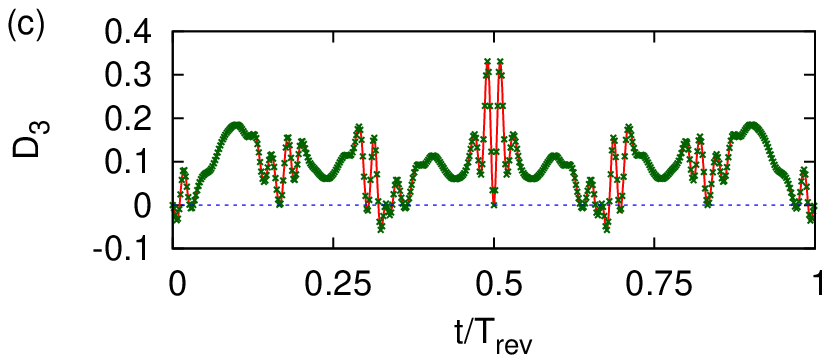}
\includegraphics[width=0.49\textwidth]{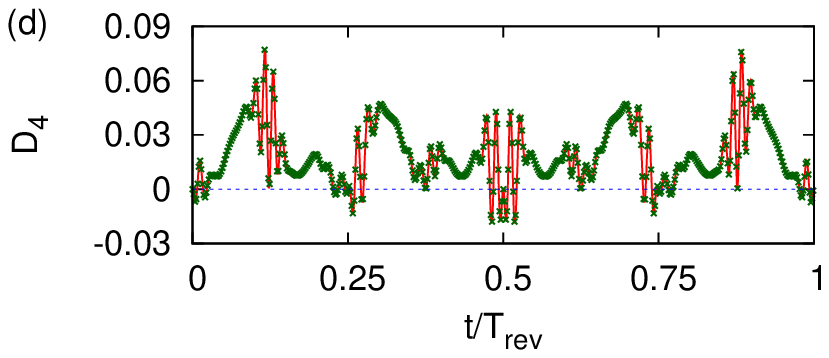}
\caption{$D_{q}(t)$ versus  $t/T_{rev}$ for $|\alpha_{a}|=|\alpha_{b}|=1$ and (a) $q=1$, (b) $q=2$, 
(c) $q=3$, and (d)$ q=4$. The squeezing parameter obtained directly from the tomogram is denoted by green crosses and that from the state by the red solid curve. The horizontal line corresponds to $D_{q}(t) = 0$.}
\label{fig:2_mode_hillery_time_dep}
\end{figure}

In figures~\ref{fig:2_mode_hillery_time_dep} (a)-(d) $D_{q}(t)$ obtained both from the states  and directly from the tomogram  for an initial state $\ket{\psi_{00}}$ are compared and seen to be in excellent agreement. It is clear from the figures that for higher values of $q$ there are more instants of time when higher-order squeezing occurs as expected from the fact that more cross terms involving the creation and destruction operators arise with increase in $q$. 
We have also verified that  both the Hong-Mandel and Hillery type squeezing and higher-order squeezing parameters obtained from tomograms  and from expectation values of appropriate operators  for the initial states $\ket{\psi_{10}}$ and $\ket{\psi_{11}}$ were equal to each other at all instants between $t =0$ and $T_{rev}$.  

\subsection{Subsystem entropies from tomograms}
This bipartite system provides an ideal framework for investigating a subsystem's nonclassical properties from the tomogram. We are primarily interested in computing the quantum information entropy and entropic squeezing properties of the subsystem as it evolves in time. Further, we extract the subsystem von Neumann entropy from the tomogram to quantify the extent of entanglement between the two condensates trapped in the double well.  (Without loss of generality we have considered subsystem $A$  by setting $\theta_{2} = 0$, and integrating over the full range of $X_{\theta_{2}}$ to obtain $w(X_{\theta_{1}}, \theta_{1})$).  

\begin{figure}
\centering
\includegraphics[width=0.49\textwidth]{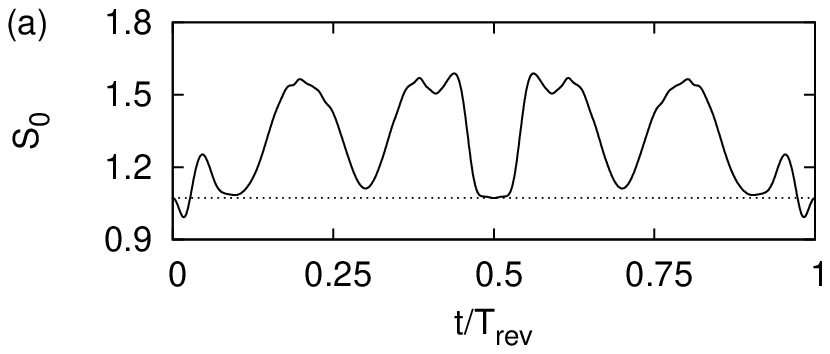}
\includegraphics[width=0.49\textwidth]{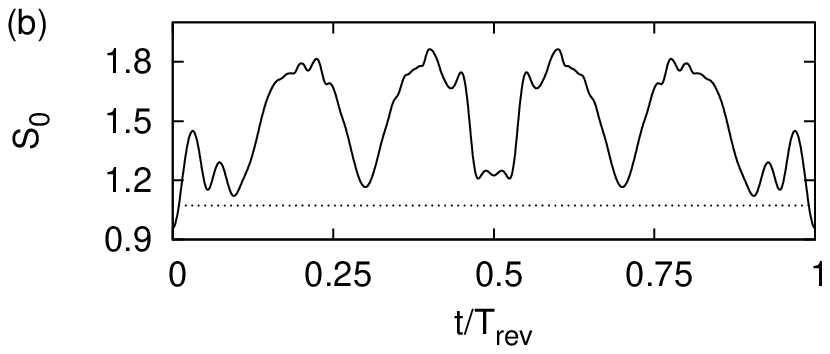}
\includegraphics[width=0.49\textwidth]{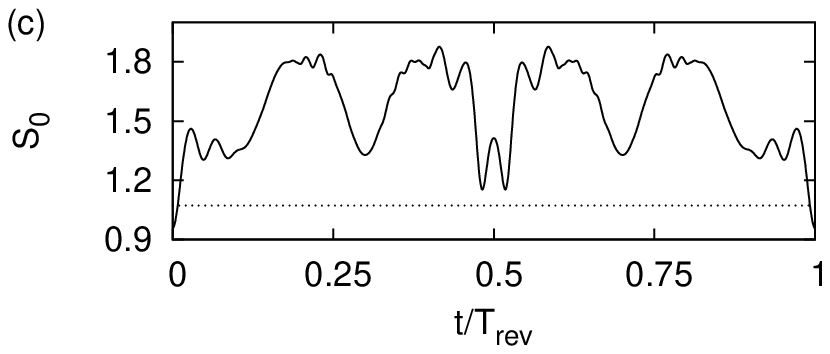}
\caption{Quantum information entropy $S_{0}$ versus $t/T_{rev}$ for initial states (a) $\ket{\psi_{00}}$, (b) $\ket{\psi_{10}}$, and (c) $\ket{\psi_{11}}$ for $\alpha_{a} = \alpha_{b} = 1$.}
\label{fig:2_mode_entropy_squeezing_time_dep}
\end{figure}

In order to examine  entropic squeezing we have further set   $\theta_{1} = 0$ and investigated the manner in which  $S_{0}$ (the information entropy  of subsystem $A$  given by $- \int_{-\infty}^{\infty}  \rmd X_{\theta_{1}} \,w_{1} \, \log w_{1}$) varies as the system evolves in time.  (Here $w_{1}$ denotes $w(X_{\theta_{1}})$).    
This information entropy has been  plotted in figures \ref{fig:2_mode_entropy_squeezing_time_dep} (a)-(c)  for  initial states $\ket{\psi_{00}}$,  $\ket{\psi_{10}}$ and $\ket{\psi_{11}}$ respectively.  The horizontal line in these figures denotes the numerical value below which entropic squeezing occurs in the quadrature considered.  We see that entropic squeezing occurs close to $t = 0$ and $T_{rev}$.   At other instants the entropy is significantly higher in the case of initial states $\ket{\psi_{10}}$ and $\ket{\psi_{11}}$  compared to initial ideal coherence.  This feature is very prominent close to $T_{rev}/2$. Further, comparing (b) and (c) it is clear that $S_{0}$ is larger at all instants if both subsystems  depart from coherence initially,  compared to the case where one of them displays initial coherence.

\begin{figure}
\centering
\includegraphics[scale=0.7]{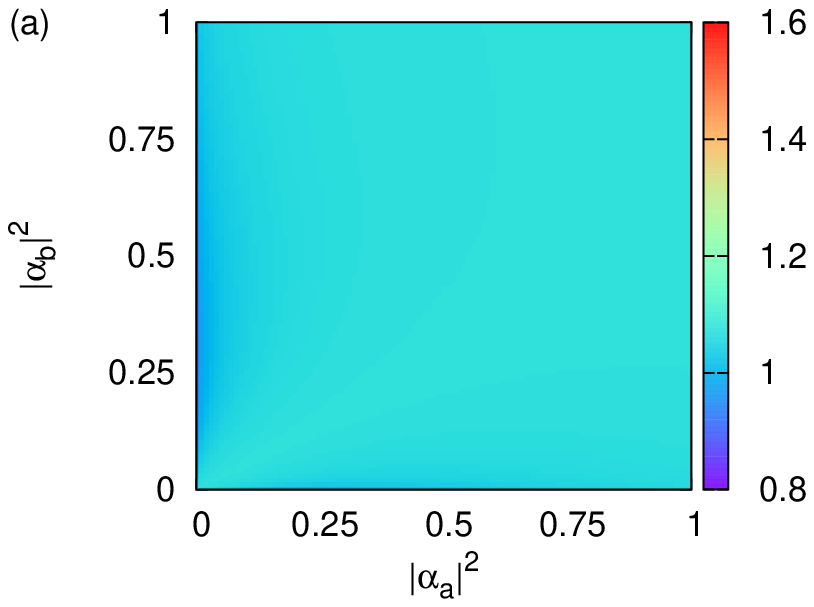}
\includegraphics[scale=0.7]{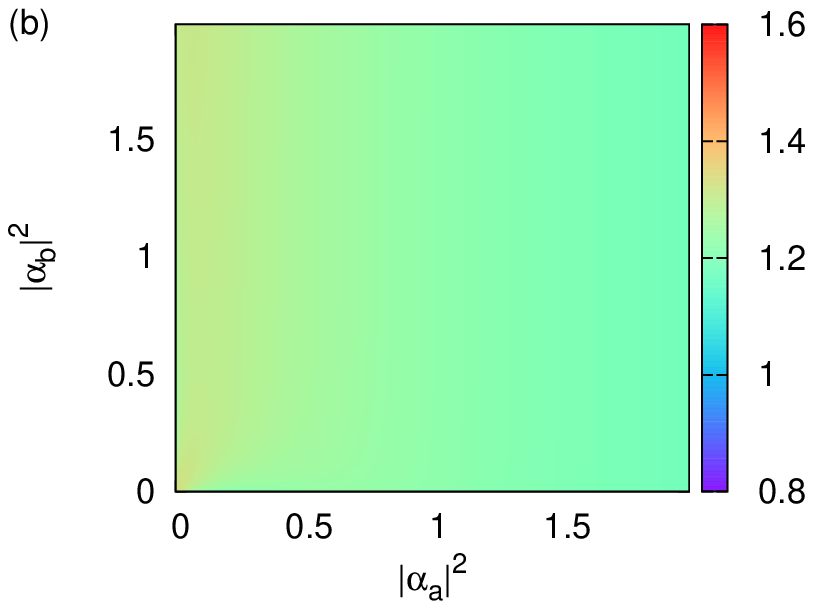}
\includegraphics[scale=0.7]{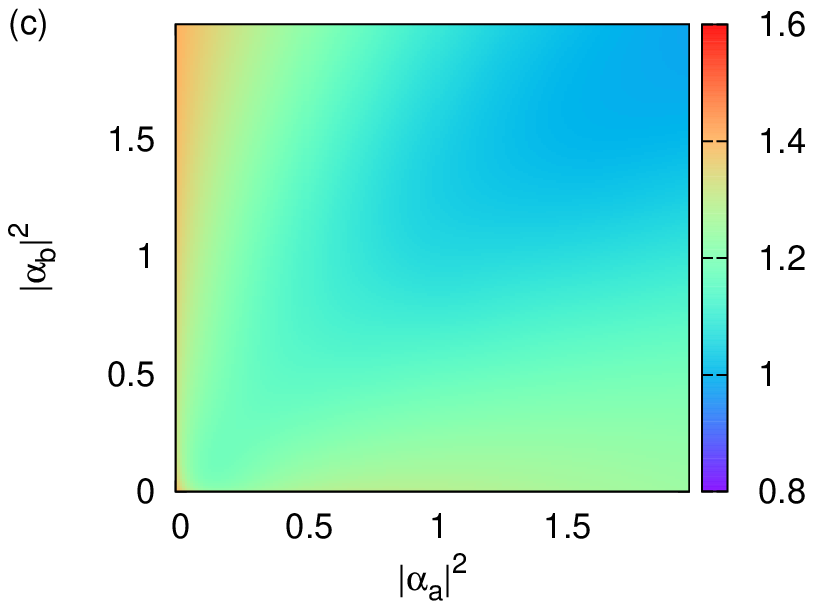}
\includegraphics[scale=0.7]{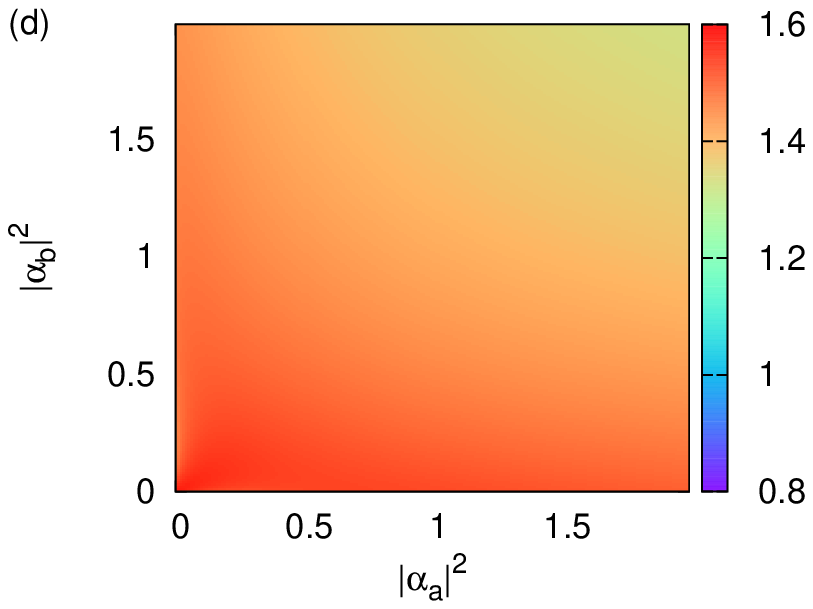}
\caption{Quantum information entropy $S_{0}$  versus $|\alpha_{a}|^{2}$ and $|\alpha_{b} |^{2}$  at $T_{rev}/2$ for initial states (a) $\ket{\psi_{00}}$, (b) and (c) $\ket{\psi_{10}}$, and (d) $\ket{\psi_{11}}$.}
\label{fig:2_mode_entropy_squeezing_alpha_dep}
\end{figure}

Figures \ref{fig:2_mode_entropy_squeezing_alpha_dep} (a), (b) and (d) show the variation of $S_{0}$  corresponding to subsystem $A$  with $|\alpha_{a}|^{2}$ and $|\alpha_{b}|^{2}$,  at $T_{rev}/2$,  for initial states $\ket{\psi_{00}}$, $\ket{\psi_{10}}$ and $\ket{\psi_{11}}$ respectively. \Fref{fig:2_mode_entropy_squeezing_alpha_dep} (c) corresponds to the entropy of subsystem $B$ for an initial state $\ket{\psi_{10}}$. This facilitates comparison of the features in figures~\ref{fig:2_mode_entropy_squeezing_alpha_dep} (b) and (c)  where for the same asymmetric initial state the two subsystems examined are different. It is evident that $S_{0}$ corresponding to $A$ is not squeezed while that corresponding to $B$ exhibits squeezing for some values of $\alpha_{a}$ and $\alpha_{b}$. The role played by the asymmetry in the initial states of the two subsystems is thus clearly brought out in these figures. It is also clear from figures \ref{fig:2_mode_entropy_squeezing_alpha_dep} (a)-(d) that entropic squeezing is more if the initial states of the subsystems are coherent.

Finally, we identify the extent of entanglement between $A$ and $B$  in terms of quantities which are  accessible from the tomogram. `Information transfer'  between $A$ and $B$ takes place because of the interaction and consequent entanglement between them. Hence we investigate the possibility of defining a quantum analogue of mutual information between two classical systems which could be used as an entanglement measure.
We recall that $w_{1}$ is independent of the value of $\theta_{2}$ and hence the information entropy
\begin{equation}
S(\theta_{1})= - \int_{-\infty}^{\infty}  \rmd X_{\theta_{1}} \,w_{1} \, \log w_{1},
\end{equation}
is also independent of the value of $\theta_{2}$. A similar statement holds for $S(\theta_{2})$. 
(Recall that $S_{0}$ is merely $S_{\theta_{1} = 0}$.)
However, 

\begin{eqnarray}
\nonumber S(\theta_{1},\theta_{2}) = - \int_{-\infty}^{\infty}  \rmd X_{\theta_{1}} &\int_{-\infty}^{\infty}  \rmd X_{\theta_{2}} \,  w(X_{\theta_{1}}, \theta_{1}; X_{\theta_{2}}, \theta_{2}) \\
&\log w(X_{\theta_{1}}, \theta_{1}; X_{\theta_{2}}, \theta_{2}).
\end{eqnarray}
clearly depends on the values of both $\theta_{1}$ and $\theta_{2}$.

We can now define, following the notation in \cite{niel}
\begin{equation}
S(\theta_{1}:\theta_{2})=S(\theta_{1})+S(\theta_{2})-S(\theta_{1},\theta_{2}).
\label{eqn:MI_analogue}
\end{equation}

$S(\theta_{1}:\theta_{2})$ is analogous to the classical mutual information.
 We can now get the  entanglement measure 
 $S(A:B)$ by averaging $S(\theta_{1}:\theta_{2})$ over $\theta_{1}$ and $\theta_{2}$.  We have numerically verified that with as few as five values of $\theta_{i}$ ($i = 1,2$),  which are equally spaced over  the interval $[0,\pi]$ and hence $25$ different pairs ($\theta_{1}$, $\theta_{2}$),  the temporal behavior of  $S(A:B)$ effectively mimics that of the SVNE or the SLE.

\begin{figure*}
\includegraphics[width=0.32\textwidth]{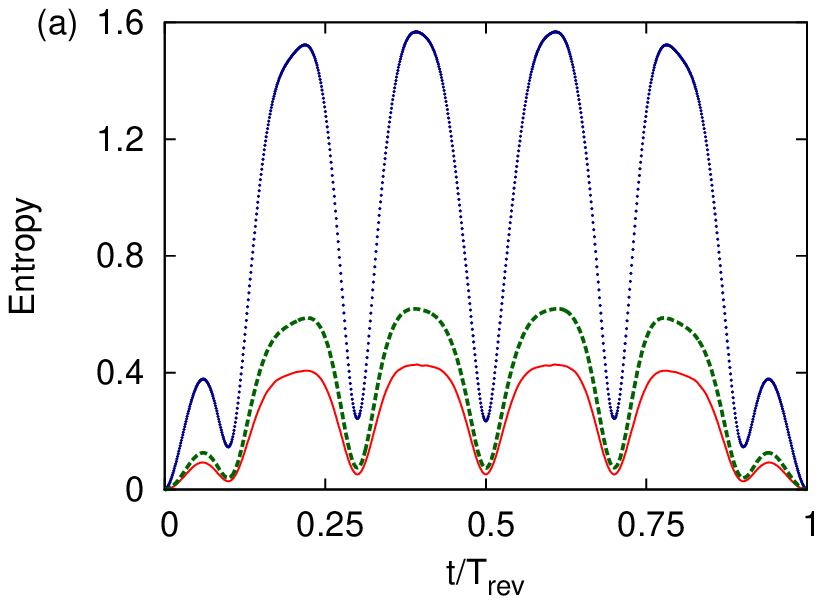}
\includegraphics[width=0.32\textwidth]{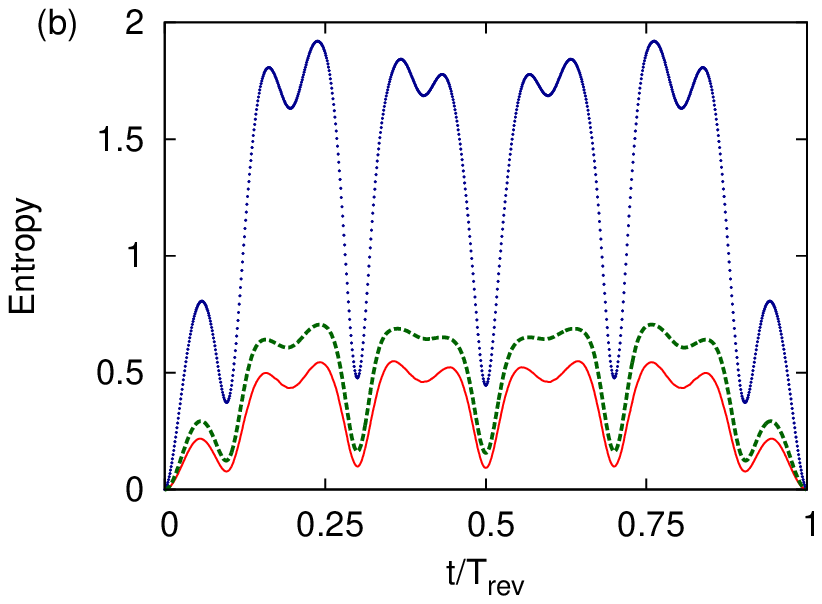}
\includegraphics[width=0.32\textwidth]{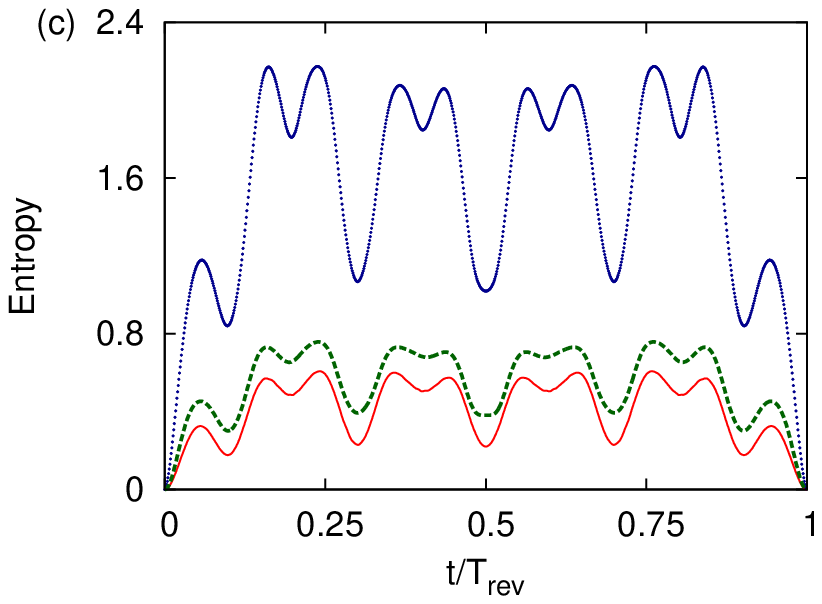}
\caption{Entanglement measures versus  $t/T_{rev}$ for initial states  (a)$\ket{\psi_{00}}$, (b) $\ket{\psi_{10}}$ and (c) $\ket{\psi_{11}}$ with $\alpha_{a} = \alpha_{b} = 1$.   $S(A:B)$,  SVNE and SLE are respectively given by the solid red curve, the blue dotted curve and the green dashed curve.}
\label{fig:2_mode_SVNE}
\end{figure*}

This is evident from figures \ref{fig:2_mode_SVNE} (a)-(c) where we have set $\alpha_{a} = \alpha_{b} = 1$ and considered  the three initial states $\ket{\psi_{00}}$, $\ket{\psi_{10}}$ and $\ket{\psi_{11}}$.
 
 In this paper we have established how tomograms can be exploited to identify and characterize a variety of nonclassical effects such as the wave packet revival phenomena and squeezing and higher-order squeezing in both single-mode and bipartite systems. While a simple relation has been shown to exist between the number of strands in tomogram patterns  and the nature of fractional revivals when a single-mode radiation field propagates in a Kerr-like medium \cite{sudhrohithrev} our investigations reveal that this no longer holds even in single-mode systems which display super-revivals during temporal evolution. We have also analyzed the revival phenomena in bipartite systems such as the double-well BEC evolving in time, solely from tomograms. 
We have obtained the extent of squeezing and also higher-order Hong-Mandel and Hillery type squeezing  for both systems from tomograms. 
  We have also investigated entropic squeezing and the extent of bipartite entanglement in detail from the tomograms alone in the case of the double-well BEC and suggested how an entanglement measure which mimics the SVNE and the SLE can be obtained from the tomograms at all instants of time. In the single-mode example we have considered initial states which are either CS or PACS and in the bipartite system the initial states are factored products of CS and states which marginally depart from macroscopic coherence.
 
 We have also undertaken similar investigations on the  Jaynes-Cummings model of field-atom interactions. This helps in identifying how tomograms  of systems with a few atomic energy levels  differ from those of multi-level atoms. These results will be reported elsewhere. 

 The primary advantage of studying experimentally obtained tomograms in detail is that the entire state-reconstruction machinery can be bypassed in examining the nature and extent of nonclassical effects. This tomographic approach becomes particularly useful when examining physical systems which are evolving in time, because state reconstruction procedures at several different instants can be avoided. We have established how a wide spectrum of nonclassical effects can be quantitatively evaluated solely from tomograms.
\appendix
\section*{Appendix}
\setcounter{section}{1}
\subsection*{Numerical computation of the two-mode density matrix}
We outline the essential steps in computing the density matrix $\rho_{m_{1},m_{2}}(t)$ of the double-well BEC, for initial states  $\ket{\psi_{m_{1}m_{2}}} = \ket{\alpha_{a},m_{1}} \otimes \ket{\alpha_{b},m_{2}}$ with Hamiltonian $H_{bec}$ \eref{eqn:2_mode_Hamiltonian}. 

The procedure for obtaining $\rho_{0,0}(t)$ the time-evolved density matrix corresponding to the initial state $\ket{\alpha_{a}} \otimes \ket{\alpha_{b}}$ is outlined in \cite{sanz}.  We obtain  $\rho_{m_{1},m_{2}}(t)=\ket{\psi_{m_{1}m_{2}}(t)}\bra{\psi_{m_{1}m_{2}}(t)}$ from $\rho_{0,0}(t)$ through appropriate transformations.  
We first write
\begin{equation}
\ket{\psi_{m_{1}m_{2}}(t)}=M_{m_{1},m_{2}}(t) \ket{\psi_{00}(t)},
\end{equation} 
where
\begin{equation}
M_{m_{1},m_{2}}(t)=\frac{1}{\kappa}\exp(-\rmi H_{bec} t) a^{\dagger m_{1}} b^{\dagger m_{2}} \exp(\rmi H_{bec} t), 
\end{equation}
and $\kappa$ is given in terms of Laguerre polynomials as $\sqrt{m_{1}! L_{m_{1}}(-{|\alpha_{a}|}^{2}) m_{2}! L_{m_{2}}(-{|\alpha_{b}|}^{2})}$.  In order to recast $M_{m_{1},m_{2}}(t)$ in a simpler form we introduce the operator $V=\exp(\gamma (a^{\dagger}b-b^{\dagger}a)/2)$, where $\gamma=\cos^{-1}(\omega_{1}/\lambda_{1})$\cite{sanz}. Consequently, $H_{bec}$ can be written as  $V H_{V} V^{\dagger}$, where $H_{V}=\omega_{0} N_{tot} + \lambda_{1} (a^{\dagger}a - b^{\dagger}b) + U_{ab} N_{tot}^{2}$.  Hence,
\begin{eqnarray}
\nonumber M_{m_{1},m_{2}}(t)=\frac{1}{\kappa} V \exp(-\rmi H_{V} t) V^{\dagger}& a^{\dagger m_{1}} b^{\dagger m_{2}}\\
& V \exp(\rmi H_{V} t) V^{\dagger}.
\end{eqnarray}
This expression can now be simplified using the following identities which can be obtained in a straightforward manner by using the Baker-Hausdorff lemma.
\begin{eqnarray}
V^{\dagger} a^{\dagger} V = a^{\dagger} \cos(\gamma / 2) + b^{\dagger} \sin(\gamma / 2),\\
V^{\dagger} b^{\dagger} V = b^{\dagger} \cos(\gamma / 2) - a^{\dagger} \sin(\gamma / 2),\\
V a^{\dagger} V^{\dagger} = a^{\dagger} \cos(\gamma / 2) - b^{\dagger} \sin(\gamma / 2),\\
V b^{\dagger} V^{\dagger} = b^{\dagger} \cos(\gamma / 2) + a^{\dagger} \sin(\gamma / 2),\\
\nonumber \exp(-\rmi \lambda_{1} (a^{\dagger}a -b^{\dagger}b) t) a^{\dagger p} b^{\dagger q} \exp(\rmi \lambda_{1} (a^{\dagger}a -b^{\dagger}b) t)\\
= a^{\dagger p} b^{\dagger q} \exp(-\rmi (p-q) \lambda_{1} t), \\
\nonumber \exp(-\rmi \omega_{0} N_{tot} t) a^{\dagger p} b^{\dagger q} \exp(\rmi \omega_{0} N_{tot} t)\\
= a^{\dagger p} b^{\dagger q} \exp(-\rmi (p+q) \omega_{0} t),\\
\nonumber \exp(-\rmi U_{ab} N_{tot}^{2} t) a^{\dagger p} b^{\dagger q} \exp(\rmi U_{ab} N_{tot}^{2} t) \\
=a^{\dagger p} b^{\dagger q} \exp(-\rmi U_{ab} t (p+q) (2 N_{tot} + p + q)).
\end{eqnarray}
Further, using binomial expansions for the two commuting operators $a^{\dagger}$ and $b^{\dagger}$ and defining $p_{max}=(k+m_{2}-l)$ and $q_{max}=(l+m_{1}-k)$, we arrive at the following simplified expression for $M_{m_{1},m_{2}}(t)$.
\begin{eqnarray}
\nonumber &M_{m_{1},m_{2}}(t)= \frac{1}{\kappa} \biggl[\sum_{k=0}^{m_{1}}\sum_{l=0}^{m_{2}}\sum_{p=0}^{p_{max}}\sum_{q=0}^{q_{max}}(-1)^{k-p} {m_{1} \choose k} {m_{2} \choose l} \\
\nonumber &\hspace{2 em} {p_{max} \choose p} {q_{max} \choose q} \exp(-\rmi\lambda_{1} t (2(k-l)+m_{2}-m_{1}))\\ 
\nonumber &\hspace{2 em} (\cos(\gamma/2))^{(k+l+p+q)} (\sin(\gamma/2))^{(2(m_{1}+m_{2})-(k+l+p+q))}\\
\nonumber &\hspace{2 em} a^{\dagger (p+q_{max}-q)} b^{\dagger (q+p_{max}-p)} \biggr] \exp(-\rmi\omega_{0} t(m_{1}+m_{2}))\\
& \times \exp(-\rmi U_{ab} t (m_{1}+m_{2}) (2 N_{tot} + m_{1}+m_{2})).
\label{eqn:appen_intermed_rho_numerics}
\end{eqnarray}
The density matrix can now be expressed in terms of $M_{m_{1},m_{2}}(t)$ and $ \rho_{0,0} (t)$ as
\begin{equation}
\rho_{m_{1},m_{2}} (t) = M_{m_{1},m_{2}}(t) \rho_{0,0}(t) M^{\dagger}_{m_{1},m_{2}}(t).
\label{eqn:appen_rho_numerics}
\end{equation}


\section*{References}
\bibliography{references}

\end{document}